\documentclass[aps,prc,superscriptaddress,reprint,nofootinbib]{revtex4-1}
\pdfoutput=1

\usepackage{epsfig}
\usepackage{dcolumn}
\usepackage{bm}
\usepackage{amssymb}
\usepackage{amsmath}
\usepackage[colorlinks=true,linktocpage=true,linkcolor=blue,citecolor=blue,allcolors=blue]{hyperref}

\usepackage{graphicx}
\usepackage[dvipsnames]{xcolor}
\usepackage{color}
\definecolor{darkblue}{rgb}{0.0, 0.0, 0.55}
\definecolor{cite}{rgb}{0.0, 0.34, 0.25}
\definecolor{midgreen}{rgb}{0.52, 0.73, 0.4}

\usepackage{rotating}
\usepackage{bigints}

\begin{document}
\title{Constraining the hadronic spectrum and repulsive interactions in a hadron resonance gas via fluctuations of conserved charges}

\author{J. M. Karthein}
\email[Corresponding author: ]{jamie@karthein.com}
\affiliation{Department of Physics, University of Houston, Houston, TX 77204, U.S.A.}
\affiliation{Nuclear Science Division, Lawrence Berkeley National Laboratory, 1 Cyclotron Road, Berkeley, CA 94720, U.S.A.}

\author{V. Koch}
\affiliation{Nuclear Science Division, Lawrence Berkeley National Laboratory, 1 Cyclotron Road, Berkeley, CA 94720, U.S.A.}

\author{C. Ratti}
\affiliation{Department of Physics, University of Houston, Houston, TX 77204, U.S.A.}

\author{V. Vovchenko}
\affiliation{Nuclear Science Division, Lawrence Berkeley National Laboratory, 1 Cyclotron Road, Berkeley, CA 94720, U.S.A.}

\date{\today}

\begin{abstract}
We simultaneously incorporate two common extensions of the hadron resonance gas model, namely the addition of extra, unconfirmed resonances to the particle list and the excluded volume repulsive interactions. We emphasize the complementary nature of these two extensions and identify combinations of conserved charge susceptibilities that allow to constrain them separately. In particular, ratios of second-order susceptibilities like $\chi_{11}^{BQ}/\chi_2^B$ and $\chi_{11}^{BS}/\chi_2^B$ are sensitive only to the baryon spectrum, while fourth-to-second order ratios like $\chi_4^B/\chi_2^B$, $\chi_{31}^{BS}/\chi_{11}^{BS}$, or $\chi_{31}^{BQ}/\chi_{11}^{BQ}$ are mainly determined by repulsive interactions. Analysis of the available lattice results suggests the presence of both the extra states in the baryon-strangeness sector and the repulsive baryonic interaction, with indications that hyperons have a smaller repulsive core than non-strange baryons. The modified hadron resonance gas model presented here significantly improves the description of lattice QCD susceptibilities at chemical freeze-out and can be used for the analysis of event-by-event fluctuations in heavy-ion collisions.

\end{abstract}

\maketitle

\section{Introduction}
\label{sec:intro}
Significant theoretical and experimental effort is dedicated to mapping out the QCD phase diagram in the temperature $T$ and baryon chemical potential $\mu_B$ plane, and to search for the elusive critical point \cite{Aoki:2006we,Aoki:2006br,Borsanyi:2010bp,Bazavov:2011nk,Borsanyi:2013bia,Bazavov:2014pvz,Fodor:2018wul,Busza:2018rrf,Stephanov:2011pb,Stephanov:1998dy,Ratti:2006gh,Fukushima:2010bq,Critelli:2017oub,Luo:2017faz,Parotto:2018pwx,Grefa:2021qvt,Karthein:2021nxe,Adamczyk:2017iwn,Adamczewski-Musch:2020slf,Abelev:2013vea,Fu:2021oaw} (for recent reviews see e.g. \cite{Ding:2015ona,Ratti:2018ksb,Bzdak:2019pkr,Ratti:2021ubw}). Ultrarelativistic heavy-ion collisions are generating the deconfined phase of strongly interacting matter in the laboratory. By systematically decreasing the collision energy of the incoming nuclei, the Relativistic Heavy Ion Collider (RHIC) at Brookhaven National Laboratory is scanning the phase diagram in the so-called Second Beam Energy Scan (BESII), soon to be followed by even lower collision energies at NICA and at the GSI-FAIR accelerator.

First-principles lattice QCD simulations are available for several thermodynamic quantities, such as the equation of state at zero \cite{Borsanyi:2010cj,Borsanyi:2013bia,Bazavov:2014pvz} and small chemical potential \cite{Guenther:2017hnx,Gunther:2017sxn,Bazavov:2017dus,Borsanyi:2021sxv,Mondal:2021jxk}, QCD transition line \cite{Bellwied:2015rza,Bazavov:2018mes,Borsanyi:2020fev,Borsanyi:2021sxv}, as well as diagonal \cite{Borsanyi:2011sw,Borsanyi:2013hza,Bellwied:2013cta,Borsanyi:2014ewa,Bellwied:2015lba,Noronha-Hostler:2016rpd,Bazavov:2017tot,Borsanyi:2018grb,Bazavov:2020bjn} and off-diagonal \cite{Borsanyi:2018grb} fluctuations of conserved charges. However, they are currently limited to the low baryonic chemical potential regime, due to the fermionic sign problem. Effective models that can reproduce lattice QCD results in certain regimes of temperature and chemical potential are therefore very useful to extend the coverage of the phase diagram beyond the reach of lattice QCD.
The models are also necessary to make a connection to the common heavy-ion observables, such as the measurements of various distributions of identified hadrons.

In the low-temperature regime ($T \lesssim 160$), the bulk thermodynamics of QCD is generally well described by a multi-component gas of free hadrons and resonances \cite{Bazavov:2012jq,Borsanyi:2010bp,Vovchenko:2016rkn,Alba:2017mqu} the so-called Hadron Resonance Gas (HRG) model. This indicates that hadron interactions in this regime may be dominated by the formation of known resonances. The HRG model has been widely used to study the confined phase of QCD matter below the transition line \cite{Karsch:2003vd,Karsch:2003zq,Tawfik:2004sw,Huovinen:2009yb,Ratti:2010kj,Alba:2015iva,Huovinen:2017ogf,Bellwied:2019pxh,Vovchenko:2014pka}. Its remarkable agreement with the equation of state from first-principles lattice calculations has led to its popularity, especially in the study of the chemical freeze-out in HICs \cite{Becattini:2012xb,Alba:2014eba,Vovchenko:2015idt,Andronic:2017pug,Bellwied:2018tkc,Alba:2020jir,Bluhm:2018aei,Flor:2020fdw}. 
However, with the availability of more differential observables like susceptibilities of conserved charges, discrepancies between the predictions of the fundamental theory and the HRG model have been observed \cite{Bazavov:2014xya,Borsanyi:2018grb,Bellwied:2021nrt}, specifically at temperatures $T \sim 150-160$~MeV that characterize the chemical freeze-out in heavy-ion collisions at the highest energies. In particular, lattice QCD results for the partial pressures have shown a need for more resonances in the strange sector than those which are already experimentally well known \cite{Bazavov:2014xya,Alba:2017mqu}. On the other hand, some susceptibility ratios including $\chi_4^B/\chi_2^B$ and $\chi_4^S/\chi_2^S$ suggest the need for repulsive interactions \cite{Bellwied:2015lba,Bazavov:2018mes,Alba:2017mqu}.

Several extensions of the HRG model have thus been proposed to improve the agreement with lattice QCD.
One possibility is extending the hadronic spectrum -- the model input -- to include more states not yet observed \cite{Bazavov:2014xya,Alba:2017mqu,Alba:2020jir}.
Other extensions incorporate additional, non-resonant interactions between hadrons such as excluded volume~\cite{Yen:1997rv,Andronic:2012ut,NoronhaHostler:2012ug,Bhattacharyya:2013oya,Vovchenko:2014pka,Albright:2015uua,Satarov:2016peb,Vovchenko:2017xad,Alba:2017bbr,Vovchenko:2018eod,Motornenko:2020yme}, van der Waals~\cite{Vovchenko:2016rkn,Vovchenko:2017zpj,Samanta:2017yhh,Sarkar:2018mbk,Vovchenko:2020lju}, mean field~\cite{Huovinen:2017ogf,Steinert:2018zni}, or are based on scattering phase shifts~\cite{Venugopalan:1992hy,Friman:2015zua,Vovchenko:2017drx,Lo:2017lym,Dash:2018can,Fernandez-Ramirez:2018vzu,Dash:2018mep}. 
While one of the advantages of the standard HRG model is certainly the lack of free parameters, apart from the uncertainties in the hadronic spectrum, introducing additional interactions unavoidably leads to new free parameters that need to be constrained through comparison with lattice results. We propose a combination of these different corrections to the standard HRG model and investigate its agreement with several lattice results on the susceptibilities. 

In this manuscript, we consider two HRG model extensions: the excluded volume interaction in the baryon sector, and the inclusion of additional particles in the hadronic list, beyond those experimentally observed. We emphasize the complementary nature of these two extensions and identify combinations of susceptibilities of conserved charges that allow to constrain them separately. The resulting HRG model considerably improves the description of lattice QCD results, in particular those for fluctuations of conserved charges.
The model can thus be useful for the analysis of freeze-out in heavy-ion collisions, in particular event-by-event fluctuations.

\section{The HRG Model and its modifications}\label{sec:HRG_ext}
\subsection{Ideal HRG}
The partial pressure for particle species $i$ in the ideal HRG model can be written as:
\begin{equation} \label{eq:press_partitionfxn}
\centering
\begin{split}
    &P_i (T, \mu_B, \mu_Q, \mu_S) = \frac{d_i T}{2\pi^2} \int_0^{\infty} (-1)^{B_i+1} k^2 dk \, \times
    \\ &\times \ln {\left[1 + (-1)^{B_i+1} \lambda_i(T, \mu_i) \exp\left(-\frac{\sqrt{k^2 + m_i^2}}{T}\right)\right]},
\end{split}
\end{equation}
where $d_i$ is the spin degeneracy factor, $B_i$ is the baryon number of species $i$, $k$ is the momentum, $\lambda_i (T,\mu_i) = \exp[(B_i \mu_B + Q_i \mu_Q + S_i \mu_S)/T]$ is the particle fugacity, and $m_i$ is the mass of species $i$. 
The partial pressure can be presented as a series containing the modified Bessel function of the second kind by expanding the logarithm and integrating term-by-term:
\begin{equation} \label{eq:press_Bessfxn}
\centering
\begin{split}
    &P_i (T, \mu_B, \mu_Q, \mu_S) = \\
    &\frac{d_i T^2}{2\pi^2} \sum_{N=1}^{\infty} \frac{[(-1)^{B_i+1}]^{N+1} \lambda_i(T, \mu_i)}{N^2} m_i^2 K_2\left(N\frac{m_i}{T}\right).
\end{split}
\end{equation}
Taking the first term in the expansion corresponds to the Boltzmann approximation.
This approximation is sufficient for our purposes and will be used throughout.
The pressure attains the following convenient form:
\begin{equation} \label{eq:press_phi}
\centering
\begin{split}
    P_i (T, \mu_B, \mu_Q, \mu_S) &= d_i \tilde{\phi}(T,m_i) \lambda_i(T,\mu_i), \\
    \tilde{\phi}(T,m_i) &=  \frac{m_i^2 T^2}{2\pi^2} \, K_2(m_i/T).
\end{split}
\end{equation}
The full pressure in the ideal HRG model corresponds to the sum of partial pressures of all hadronic components.
It is convenient to group the contributions of the various hadrons in accordance with their quantum numbers.
For instance, introducing
\begin{equation} \label{eq:press_sectors}
\centering
    \tilde{\phi}(T) = \sum_{j \in \rm{sectors}} d_j \tilde{\phi}(T,m_j)
\end{equation}
allows us to identify the different sectors of the total HRG pressure broken down by various quantum numbers that are listed in Table \ref{tab:hadron_table}.
Note that all species apart from the $i = 0$ sector~(neutral particles) 
have a corresponding antiparticle sector. 
\begin{table}[h]
\centering
\begin{tabular}{c|c c c|c}
 $i$ & $B$ & $Q$ & $S$ & base hadron \\ \hline
 0 &  0 & 0  & 0 & $\pi^0$ \\ 
 1 &  0 & 1 & 0 & $\pi^+$ \\ 
 2 &  0 & 1 & 1 & K$^+$ \\ 
 3 &  0 & 0 & 1 & K$^0$ \\ 
 4 &  1 & 0 & 0 & n  \\ 
 5 &  1 & 1 & 0 & p  \\
 6 &  1 & 2 & 0 & $\Delta^{++}$  \\
 7 &  1 & -1 & 0 & $\Delta^{-}$ \\
 8 &  1 & 0 & -1 & $\Lambda$ \\
 9 &  1 & 1 & -1 & $\Sigma^+$ \\
 10 & 1 & -1 & -1 & $\Sigma^-$ \\
 11 & 1 & 0 & -2  & $\Xi^0$ \\
 12 & 1 &  -1 &  -2 & $\Xi^-$ \\
 13 & 1 & -1 & -3  & $\Omega^-$
\end{tabular}
\caption{
\label{tab:hadron_table}
The list of hadronic quantum number sets contributing to the pressure of the ideal HRG model.
The last column identifies the lowest mass hadron representing the set of quantum numbers.
}
\end{table}
Therefore, the pressure from  Eq.~(\ref{eq:press_phi}) takes the form of a truncated relativistic expansion in fugacities:
\begin{equation} \label{eq:Pfug}
    \centering
    P (T, \mu_B, \mu_Q, \mu_S) = \tilde{\phi_0}(T) + \sum_{i \neq 0} 2 \, \tilde{\phi_i}(T) \cosh \left(\mu_i/T \right),
\end{equation}
where $\mu_i=B_i \mu_B + Q_i \mu_Q + S_i \mu_S$ is the chemical potential of the corresponding $i$th sector.
Each term in Eq.~\eqref{eq:Pfug} corresponds to the partial pressure associated with the particular set of hadronic quantum numbers.

\subsection{Extended Spectrum}
As a correction to the standard HRG model, our first extension is the incorporation of hadronic states beyond those which are well-known experimentally. When considering the partial pressures from Lattice QCD, it has been shown that the hadronic spectrum in QCD goes beyond what exists in the ordinary version of the hadronic list from the Particle Data Group (PDG) with only the most well-known states \cite{Zyla:2020zbs}. 
The PDG ranks particles by how well established they are, with a rating based on a number of stars (*). 
The **** states are those which are very well established, like e.g. nucleons or $\Delta(1232)$ resonances. 
On the opposite side, the * states are the least established ones, for example $\Delta$(1750) and other high-mass resonances. 
A previous investigation of the agreement between the partial pressures from the lattice and those obtained in the HRG model with different hadronic lists across the various quantum number sectors, found that the PDG2016+ particle list reproduced the largest number of quantities calculated on the lattice without exceeding them~\cite{Alba:2017mqu}. 
This hadronic list incorporates all states from the 2016 version of the PDG~\cite{Patrignani:2016xqp}, starting from the well-known ones, all the way down to those listed as “seen”, with a confidence rating of *. 
Another possibility is to incorporate the states predicted by the Quark Model~\cite{Capstick:1986bm,Godfrey:1985xj,Ebert:2009ub}, which includes an even larger number of hadrons than those contained in the PDG2016+.
As we can see in Eq. (\ref{eq:press_phi}), any additional states seek to increase the pressure of the system, i.e. a larger number of states will lead to a larger overall pressure.

It was recently pointed out in Ref. \cite{Bollweg:2021vqf} that some states calculated within the Quark Model from Ref. \cite{Capstick:1986bm} are overlapping with states later measured by the PDG \cite{Zyla:2020zbs}. 
In light of this, we provide an update to the QM list first published in Ref. \cite{Alba:2020jir} that removes all the duplicate states that remained as an artifact of those early Quark Model calculations \cite{listrepo}.
We note that the calculations here denoted as QM are using this updated list and that we observed a minimal difference between the old and new list when calculating the fluctuations of interest in this study.

In this study, we aim at identifying the most suitable description of the lattice data, with both the excluded volume interactions and additional hadronic states incorporated.
In order to do so, we revisit the comparison with lattice data and investigate several different hadronic lists: 
\begin{itemize}
    \item PDG2016 -- ordinary hadronic list with only the well-known states *** $-$ **** from the 2016 Particle Data Booklet;
    \item PDG2016+ -- the list containing both the established~(***$-$****) and unconfirmed~(*$-$**) states;
    \item Quark Model~(QM) -- the list which incorporates all states predicted by the Quark Model.
\end{itemize}
The latter two lists were introduced and described in detail in \cite{Alba:2017mqu, Alba:2020jir}. 
We also checked that the most recent compilation of the established states from the Particle Data Group -- the PDG2020 list -- yields negligible differences compared to the PDG2016 list, thus we retain the latter list in the analysis for consistency with Refs.~\cite{Alba:2017mqu,Alba:2020jir}.
\subsection{Excluded Volume}

The next extension to the HRG model is the excluded volume model. This corresponds to including repulsive interactions between hadrons. Many versions of the EV-HRG model have been considered in the literature. Here we follow the approach introduced in Refs.~\cite{Vovchenko:2016rkn,Vovchenko:2017xad} where EV interactions are included only for baryon-baryon and antibaryon-antibaryon pairs. 
This corresponds to a minimalistic EV extension that does not affect meson-meson and meson-baryon interactions, which are presumed to be dominated by resonance formation and thus already included in the HRG model.
The pressure is partitioned into contributions of non-interacting mesons and interacting baryons and antibaryons:
\begin{equation} \label{eq:pevhrg}
\centering
p = p_M^{\rm id} + p_B^{\rm ev} + p_{\bar{B}}^{\rm ev},
\end{equation}
where
\begin{equation} \label{eq:pMev}
p_{M}^{\rm id}  = \tilde{\phi}_0(T) ~ + \sum_{i \neq 0, \, i \in M} 2 \, \tilde{\phi}_i(T) \, \cosh(\mu_i/T),
\end{equation}
\begin{equation} \label{eq:pBev}
p_{B(\bar{B})}^{\rm ev}  =  \sum_{i \in B} \tilde{\phi}_i(T) \, \exp(\pm \mu_i/T) \,
\exp\left( \frac{- b \, p_{B(\bar{B})}^{\rm ev}}{T} \right)~.
\end{equation}
Here, $i \in M$ corresponds to mesons~($B_i = 0$), $i \in B$ corresponds to baryons~($B_i = 1$), $b$ is the baryon excluded volume parameter, and $\tilde{\phi}(T)$ is given in Eq. (\ref{eq:press_phi}).
Equation~\eqref{eq:pBev} can be solved in terms of the Lambert W function~\cite{NoronhaHostler:2012ug,Taradiy:2019taz}:
\begin{equation}\label{eq:pBevW}
p_{B(\bar{B})}^{\rm ev}  = \frac{T}{b} \, W[\varkappa_{B(\bar{B})}(T,\mu_B,\mu_Q,\mu_S)],~
\end{equation}
where
\begin{equation}
\varkappa_{B(\bar{B})}(T,\mu_B,\mu_Q,\mu_S) = b \, \sum_{i \in B} \tilde{\phi}_i(T) \, \exp(\pm \mu_i/T).
\end{equation}
The explicit form Eq.~\eqref{eq:pBevW} for the pressure in the EV-HRG model in terms of the Lambert W function allows us to forgo solving the transcendental equation for the pressure.

One should note that different formulations of the excluded volume HRG model exist in the literature. 
In the formulation that we use the EV interactions are introduced only for baryon-baryon and antibaryon-antibaryon pairs with a common EV parameter $b$.
This is consistent with the model used in Refs.~\cite{Vovchenko:2016rkn,Vovchenko:2017xad}.
In other EV models~\cite{Andronic:2012ut,Alba:2017bbr,Motornenko:2020yme} the repulsive interactions are introduced for \emph{all} hadron pairs, and each hadronic species may be characterized by its own value of the EV parameter.
As such, these types of EV models contain many more free parameters, while ours is a minimalistic approach.
The EV effects influence the thermodynamics differently in the latter class of EV models, and the conclusions obtained in this work within the former class of EV models do not necessarily translate.

\section{Susceptibilities}
\label{sec:susc}

The two HRG model extensions introduced above -- the excluded volume corrections and the inclusion of extra states -- are complementary to one another.
This can be seen in the following way.
Adding extra resonances can be interpreted as adding attractive interactions among hadrons that lead to the formation of these resonances.
Given the fact that all the extra states have baryon number equal to either 0 or $\pm 1$, this may correspond to meson-meson and meson-baryon interactions, but not to baryon-baryon interactions. 
On the other hand, the EV corrections considered here correspond to repulsive baryon-baryon interactions but not to any meson-meson or meson-baryon interactions. 
Therefore, the two extensions describe different physics, and thus, they can and should be considered simultaneously.

Both extensions affect the equation of state.
For instance, the inclusion of extra states increases the pressure at a given temperature and chemical potential, while the EV interactions lead to its suppression.
It can be challenging, then, to constrain the two effects separately.
In order to achieve those constraints, we study differential observables that have recently been obtained on the lattice, namely the susceptibilities of conserved charges:
\begin{equation} \label{eq:chis}
\chi^{BQS}_{lmn} = \frac{\partial^{l+m+n}(p/T^4)}{\partial(\mu_{B}/T)^{l} \, \partial(\mu_{Q}/T)^{m} \, \partial(\mu_{S}/T)^{n}}
\end{equation}

The susceptibilities in Eq.~\eqref{eq:chis} can be calculated explicitly in the EV-HRG model with extra states by utilizing Eq.~\eqref{eq:pBevW} and the known properties of the Lambert W function.
One can then construct specific combinations of susceptibilities that are mainly sensitive to either the extra states or baryon excluded volume, but not to both.

\subsection{Extra states from baryon correlators}

Firstly,  we consider the following ratios of second order susceptibilities: $$ \chi_{11}^{BQ} / \chi_2^B, \quad \chi_{11}^{BS} / \chi_2^B.$$
By calculating them explicitly one obtains
\begin{align} \label{eq:chi11BQ}
\frac{\chi_{11}^{BQ}}{\chi_2^B} & = \frac{\sum_{j \in \rm{sectors}} \, B_j \, Q_j \, \tilde{\phi}_j (T) }{\sum_{j \in \rm{sectors}} \, B_j^2 \, \tilde{\phi}_j (T)}, \\
 \label{eq:chi11BS}
\frac{\chi_{11}^{BS}}{\chi_2^B} & = \frac{\sum_{j \in \rm{sectors}} \, B_j \, S_j \, \tilde{\phi}_j (T) }{\sum_{j \in \rm{sectors}} \, B_j^2 \, \tilde{\phi}_j (T)}.
\end{align}
The excluded volume parameter $b$ cancels out in this combination of susceptibilities.\footnote{A similar cancellation has been observed for the $\chi_{11}^{BQ}/\chi_2^B$ ratio in Ref.~\cite{Vovchenko:2017uko} in the framework of the van der Waals HRG model.} 
On the other hand, the ratios are sensitive to the particle list encoded in the ``partial pressures'' $\tilde{\phi}_j$.
Therefore, these ratios can be used to constrain the hadronic spectrum. 
In particular, it follows from Eqs.~\eqref{eq:chi11BQ} and~\eqref{eq:chi11BS} that $ \chi_{11}^{BQ} / \chi_2^B$ and $ \chi_{11}^{BS} / \chi_2^B$ probe the fractions of charged baryons and hyperons, respectively, compared to all baryons.

\begin{figure*}
    \centering
    \includegraphics[width=0.48\linewidth]{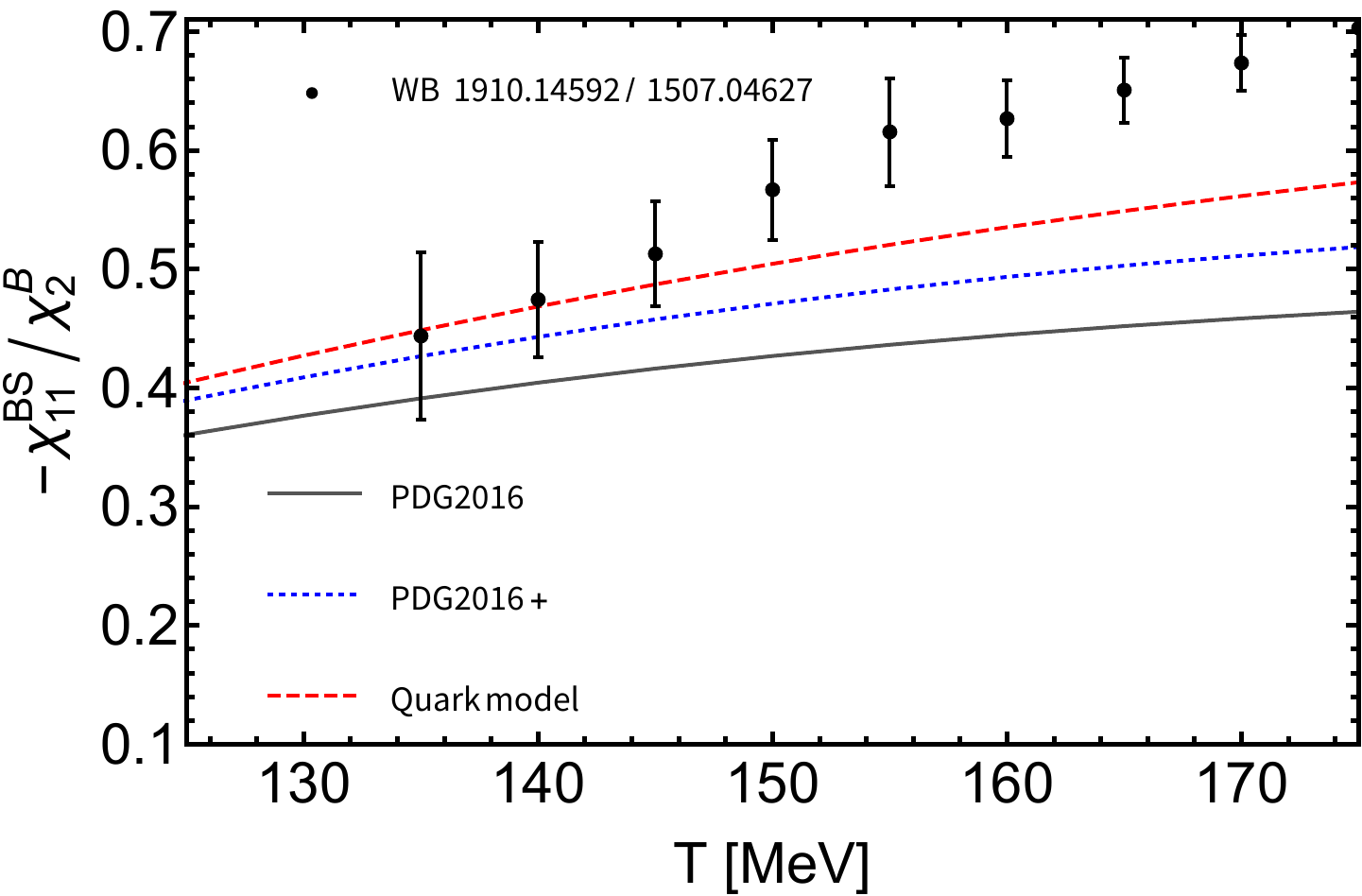}
    \includegraphics[width=0.48\linewidth]{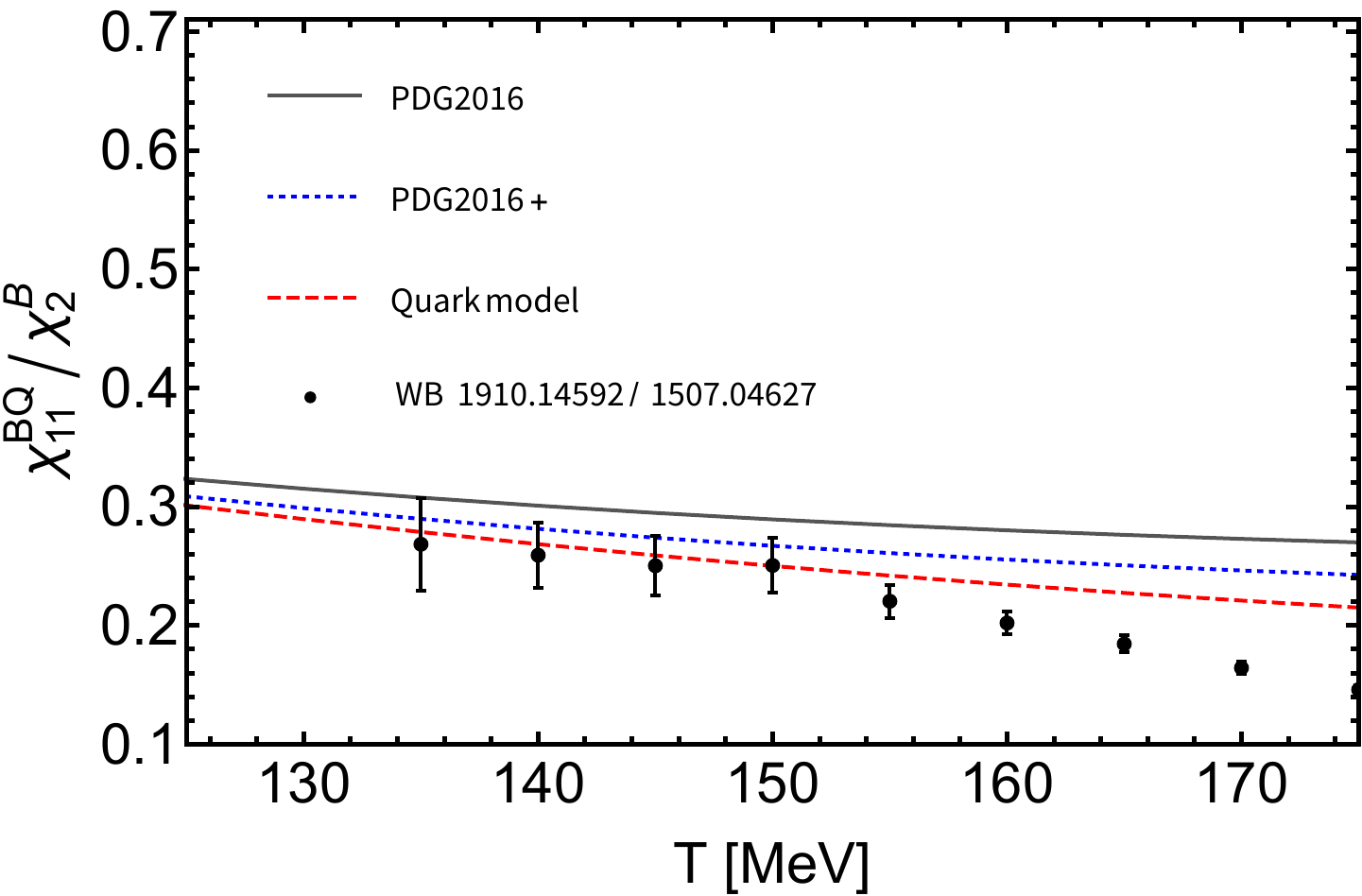}
    \caption{Temperature dependence of second order susceptibility ratios $\chi_{11}^{BS}/\chi_2^B$ and $\chi_{11}^{BQ}/\chi_2^B$. Continuum extrapolated lattice results from Refs. \cite{Bellwied:2019pxh, Bellwied:2015lba} are shown by the black points with error bars, while the  calculations within the EV-HRG model are the curves for different hadronic lists. This combination of susceptibilities leads to the cancellation of the excluded volume parameter, $b$.}
    \label{fig:chi-BS-2B_no_EV_dep}
\end{figure*}

The ratios $\chi_{11}^{BQ}/\chi_2^B$ and $\chi_{11}^{BS}/\chi_2^B$ are shown in Fig.~\ref{fig:chi-BS-2B_no_EV_dep}.
The inclusion of extra states from PDG2016+ and QM leads to the enhancement of $-\chi_{11}^{BS}/\chi_2^B$ and the suppression of $\chi_{11}^{BQ}/\chi_2^B$.
This is driven by the fact that the extra states are mainly hyperons, thus their addition increases the fraction of strange baryons~(probed by $-\chi_{11}^{BS}/\chi_2^B$) and decreases the fraction of non-strange baryons~(probed by $\chi_{11}^{BQ}/\chi_2^B$).
The comparison with continuum extrapolated lattice data \cite{Bellwied:2019pxh, Bellwied:2015lba} suggests the need for additional hyperon states from PDG2016+/QM, as previously discussed in Ref. \cite{Alba:2017mqu}. 
The best agreement with the lattice data is obtained for the QM list.

None of the considered particle lists allow us to describe the lattice data within errors at $T \gtrsim 155$~MeV.
There are several possibilities which might explain these deviations. 
A possible explanation, which would not be captured by our model or any modifications on it, is that this temperature corresponds to the onset of deconfinement, at which new degrees of freedom (quarks) start to be liberated. Otherwise, an improvement in the agreement between lattice results and the model could be obtained through one of the following considerations.
If there are even more strange baryons than predicted by the QM, this could improve the agreement with the lattice data.
For instance, the presence of broad, high-mass Hagedorn states~\cite{Hagedorn:1965st} may have a considerable effect on the susceptibilities as one approaches the Hagedorn temperature $T_H \sim 160-180$~MeV~\cite{Lo:2015cca}. 
If the Hagedorn states contain more strange baryons than non-strange baryons, this may improve the agreement with the lattice data in Fig.~\ref{fig:chi-BS-2B_no_EV_dep}.
However, it might be challenging to preserve at the same time the agreement with the individual susceptibilities rather than in the ratios alone.

Other explanations would go beyond the physics of the model employed in the present paper.
For instance, we have modeled all resonances as free particles with zero width.
On the other hand, many non-strange baryon resonances like $\Delta$'s and $N^*$'s are broad, thus a proper treatment of their spectral functions should be important.
Modeling of broad resonances is challenging, analyses in the literature based on either pion-nucleon scattering phase shifts within the S-matrix approach~\cite{Lo:2017lym} or energy-dependent Breit-Wigner widths~\cite{Vovchenko:2018fmh} indicate that partial pressures of such resonances might be overestimated in the standard HRG model.
This implies that a more involved treatment of broad resonances may lead to a suppressed $\chi_{11}^{BQ}/\chi_2^B$ ratio~(and hence an enhanced $-\chi_{11}^{BS}/\chi_2^B$) and recover the agreement with the lattice data.

Finally, the comparison with the lattice data may be affected if there is a flavor hierarchy in baryon excluded volumes. 
While the excluded volume effects cancel out in $\chi_{11}^{BQ}/\chi_2^B$ and $\chi_{11}^{BS}/\chi_2^B$ ratios when a common EV parameter $b$ is used for all baryons, this would no longer be the case if excluded volumes differ between strange and non-strange baryons.
A smaller EV for strange baryons would lead to a smaller suppression of $\chi_{11}^{BS}$ relative to $\chi_{11}^{BQ}$, thus leading to an improved agreement with the lattice data.
It is possible that a combination of the three effects discussed here is at play, and it would be interesting to study these in more detail in the future.
We indeed find indications for the flavor-dependent excluded volumes in the behavior of fourth-order susceptibilities discussed in the following subsection.

\subsection{Fourth-order cumulants and excluded volume}

In addition to the extra states, we also want to place limits on the excluded volume parameter, $b$. 
This can be done by considering ratios of fourth-to-second order susceptibilities.
The following three ratios are all equal in the EV-HRG model under consideration and sensitive to the EV parameter $b$:
\begin{align} 
\frac{\chi^B_4}{\chi^B_2} = \frac{\chi^{BS}_{31}}{\chi^{BS}_{11}} =  \frac{\chi^{BQ}_{31}}{\chi^{BQ}_{11}} 
& = \frac{1 - 8 \, W(\varkappa_B) + 6 [W(\varkappa_B)]^2}{[1 + W(\varkappa_B)]^4} \nonumber \\
& = 1 - 12 \varkappa_B + O(\varkappa_B^2).
\end{align}    
In the ideal HRG model, i.e. without the EV interactions, $\varkappa_B = 0$ so that $\frac{\chi^B_4}{\chi^B_2} = \frac{\chi^{BS}_{31}}{\chi^{BS}_{11}} = \frac{\chi^{BQ}_{31}}{\chi^{BQ}_{11}} = 1$ regardless of the inclusion of any additional hadronic states. 
The suppression of these ratios relative to unity, on the other hand, is directly sensitive to the EV interactions and can be used to constrain the EV parameter $b$.
Furthermore, the fact that all three ratios are predicted to be equal within the model allows us to probe the limits of validity of the model, which would be signaled by the point where the equality among these three ratios no longer holds in the lattice data.
\begin{figure}
    \centering
    \includegraphics[width=\linewidth]{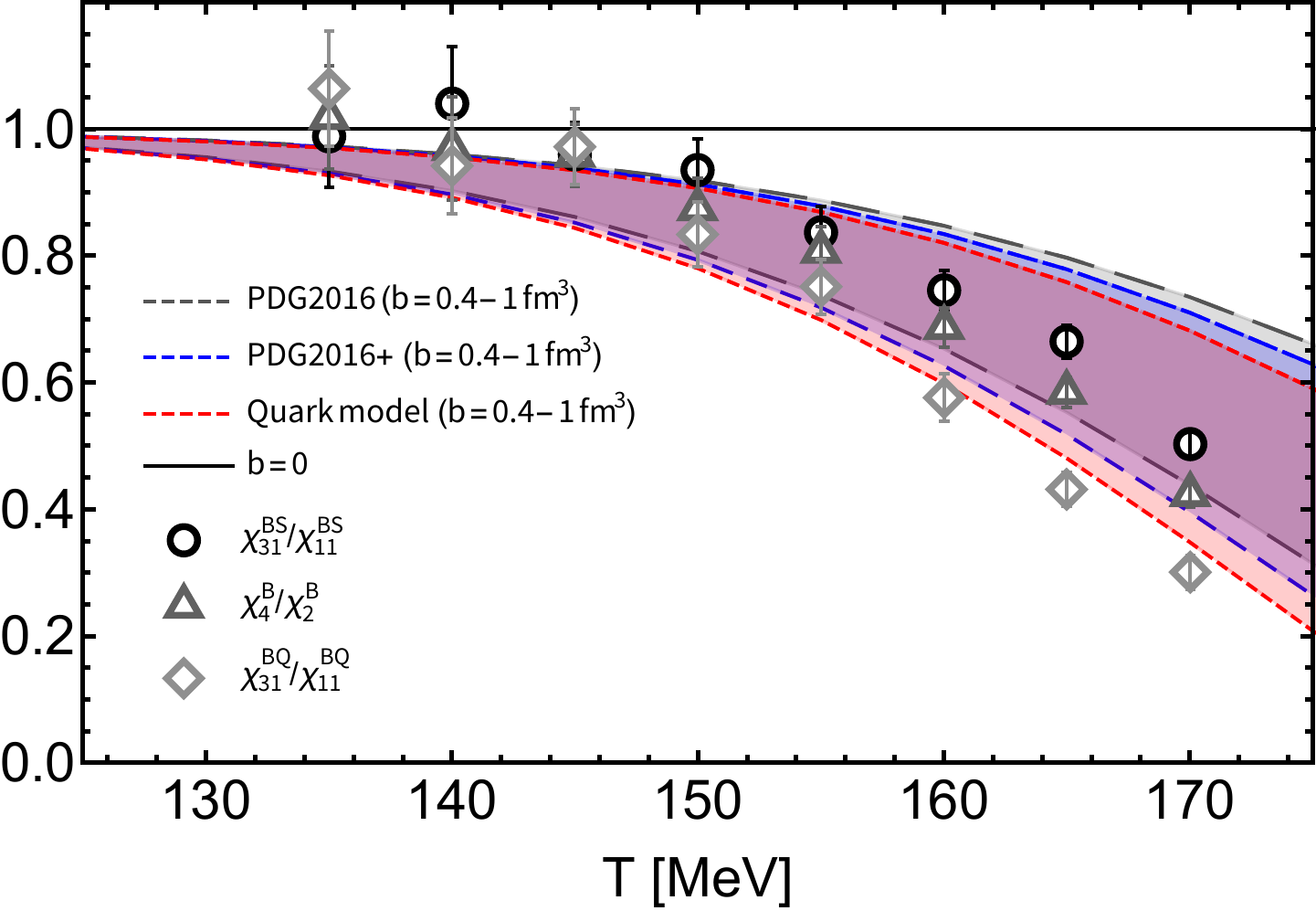}
    \caption{Temperature dependence of fourth-to-second order susceptibility ratios $\chi_4^B/\chi_2^B$, $\chi_{31}^{BS}/\chi_{11}^{BS}$, and $\chi_{31}^{BQ}/\chi_{11}^{BQ}$, predicted to be equal in the EV-HRG model. The lattice data at finite lattice spacing $N_{\tau}=12$ from Ref. \cite{Borsanyi:2018grb} are shown as grayscale symbols with error bars, while the calculations within the EV-HRG model are shown as bands for a range of excluded volume parameter, $b$, and for different hadronic lists. The ideal HRG result is given by the horizontal line at unity.
    }
    \label{fig:chi-4-2_equiv_EV}
\end{figure}

Figure~\ref{fig:chi-4-2_equiv_EV} depicts the results of the calculation of the ratios $\chi_4^B/\chi_2^B$, $\chi_{31}^{BS}/\chi_{11}^{BS}$ and $\chi_{31}^{BQ}/\chi_{11}^{BQ}$ within the EV-HRG model for range $b = 0.4-1$~fm$^3$ of the EV parameter values. 
The three ratios all coincide with one another, as expected, and exhibit
minimal dependence on the hadronic list utilized. 
Therefore, these particular quantities are indeed sensitive mainly to the excluded volume repulsive interactions rather than to the hadronic spectrum used in the HRG model. 
In Fig. \ref{fig:chi-4-2_equiv_EV} we compare the calculations in the EV-HRG model with various particle lists to lattice data at $N_{\tau} = 12$ from the Wuppertal-Budapest collaboration~\cite{Borsanyi:2018grb}, in the temperature range $T = 135-170$~MeV. 
Since not all of the available lattice data are continuum extrapolated, we choose this larger lattice spacing for the comparisons and avoid constructing ratios that would be a mixture of results at finite $N_{\tau}$ and in the continuum limit. 
The qualitative behavior of the three ratios is very similar in the whole temperature range considered.
Quantitatively, we see that at temperatures below 150 MeV the three ratios sit on top of each other. 
This is expected, although the lattice error bars are relatively sizable at those temperatures.

Statistically significant differences between the three susceptibility ratios in the lattice data emerge at $T \gtrsim$~160~MeV.
On the one hand, this may be a reflection of the transition to the deconfined phase where a hadronic model would be expected to break down.
In the Stefan-Boltzmann limit of massless quarks one has $\chi_4^B/\chi_2^B = \chi_{31}^{BS}/\chi_{11}^{BS} \to 2 / (3\pi^2)$ while $\chi_{31}^{BQ}/\chi_{11}^{BQ} \to 0$.
Thus, the notably smaller values of $\chi_{31}^{BQ}/\chi_{11}^{BQ}$ at $T \gtrsim 160$~MeV compared to the other two ratios might be related to the smaller Stefan-Boltzmann limit for this quantity.
On the other hand, these differences may also reflect a flavor hierarchy in baryon excluded volumes.
The ratios are predicted to be equal in the EV-HRG model if all baryons are assigned a common EV parameter $b$.
However, if for example strange baryons have a different~(smaller) excluded volume, one would expect $\chi_{31}^{BS}/\chi_{11}^{BS}$ to exhibit smaller deviations from the baseline of unity than the other ratios.
We see in Fig.~\ref{fig:chi-4-2_equiv_EV} that the separation between the strangeness and electric charge susceptibilities is such that they lie on the higher and lower ends of the $b = 0.4-1$~fm$^3$ band, respectively.
This is an indication that the strange baryons may indeed have a smaller volume than the non-strange ones.
From a phenomenological point of view, the smaller hyperon volume could reflect the fact that hyperon-hyperon interactions are mediated by the exchange of heavier mesons like $\phi$ compared to nucleon-nucleon interactions which correspond to the exchange of lighter, non-strange mesons like $\sigma$ and $\omega$.
The implementation of smaller excluded volumes for strange particles is possibly not unique. Such extensions of the EV-HRG model have been considered in Refs.~\cite{Alba:2016hwx,Vovchenko:2017zpj,Alba:2017bbr,Motornenko:2020yme} but are not considered in the present study. Of course it can be done in the future in order to model the subtle differences between $\chi_4^B/\chi_2^B$, $\chi_{31}^{BS}/\chi_{11}^{BS}$, and $\chi_{31}^{BQ}/\chi_{11}^{BQ}$, for example at the chemical freeze-out in heavy-ion collisions.

\subsection{Effect of the extra states and excluded volume on other susceptibilities}

Next, we investigate the following two combinations of susceptibilities that are sensitive to the extra strange states: the kurtosis of net-strangeness fluctuations $\chi_4^S / \chi_2^S$ and the correlator $\chi_{11}^{us}$ between net numbers of up and strange quarks.
The effect of extra states on these quantities was investigated in Ref.~\cite{Alba:2017mqu} without excluded volume effects. 
It was shown there that extra states improve the description of $\chi_{11}^{us}$ but in the case of the QM list spoil the agreement with the lattice data for $\chi_4^S/\chi_2^S$. 
Here we investigate how these quantities are affected by the presence of baryon excluded volume in addition to extra states.

\begin{figure}
    \centering
    \includegraphics[width=\linewidth]{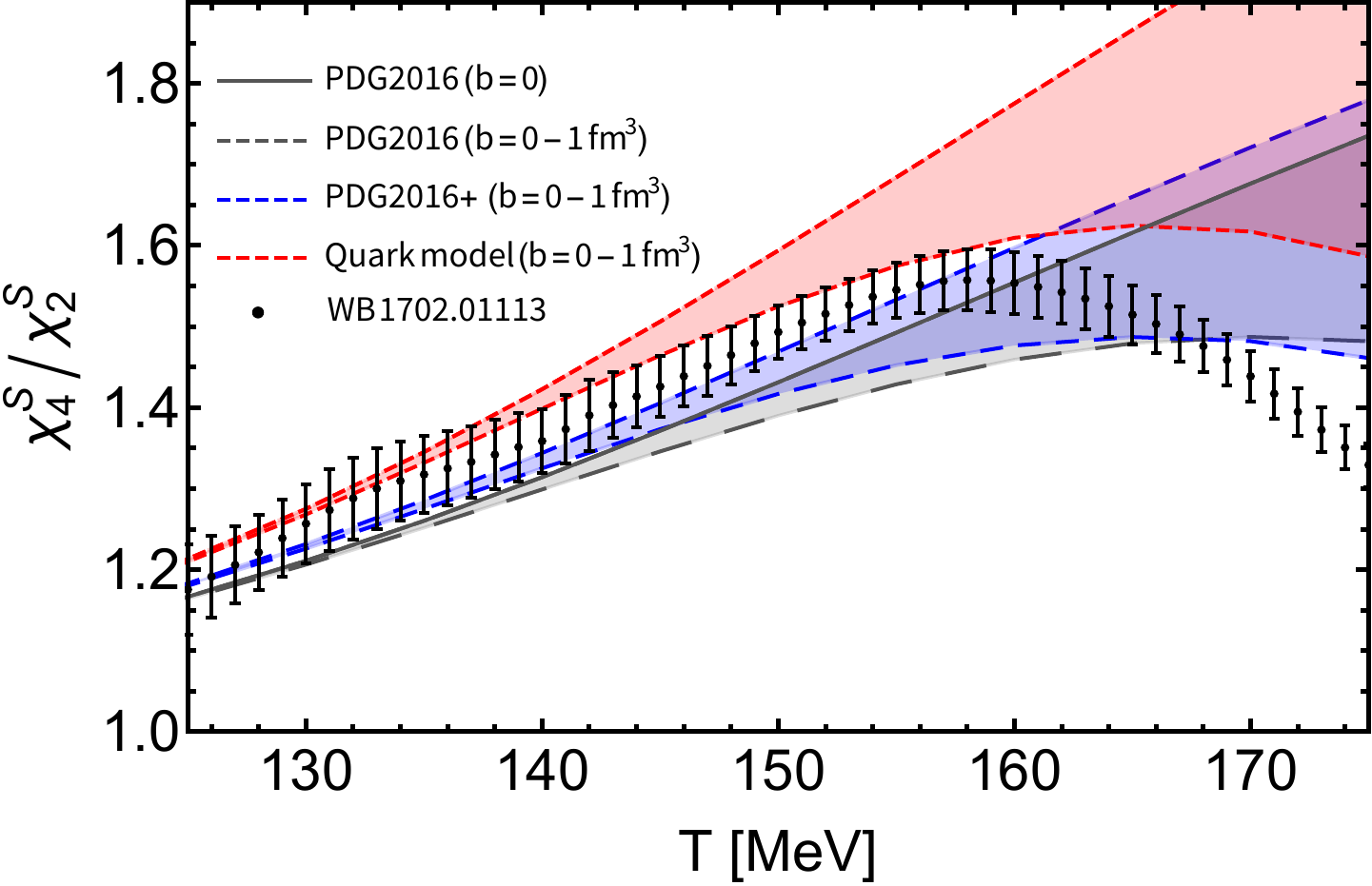}
    \caption{Temperature dependence of fourth-to-second order strangeness susceptibility ratio $\chi_4^S/\chi_2^S$. Continuum extrapolated lattice results for this quantity are given by black points with error bars \cite{Alba:2017mqu} and EV-HRG calculations are shown for the full range of the parameter $b$ for different hadronic lists.}
    \label{fig:chi-4S-2S}
\end{figure}

\begin{figure}
    \centering
    \includegraphics[width=\linewidth]{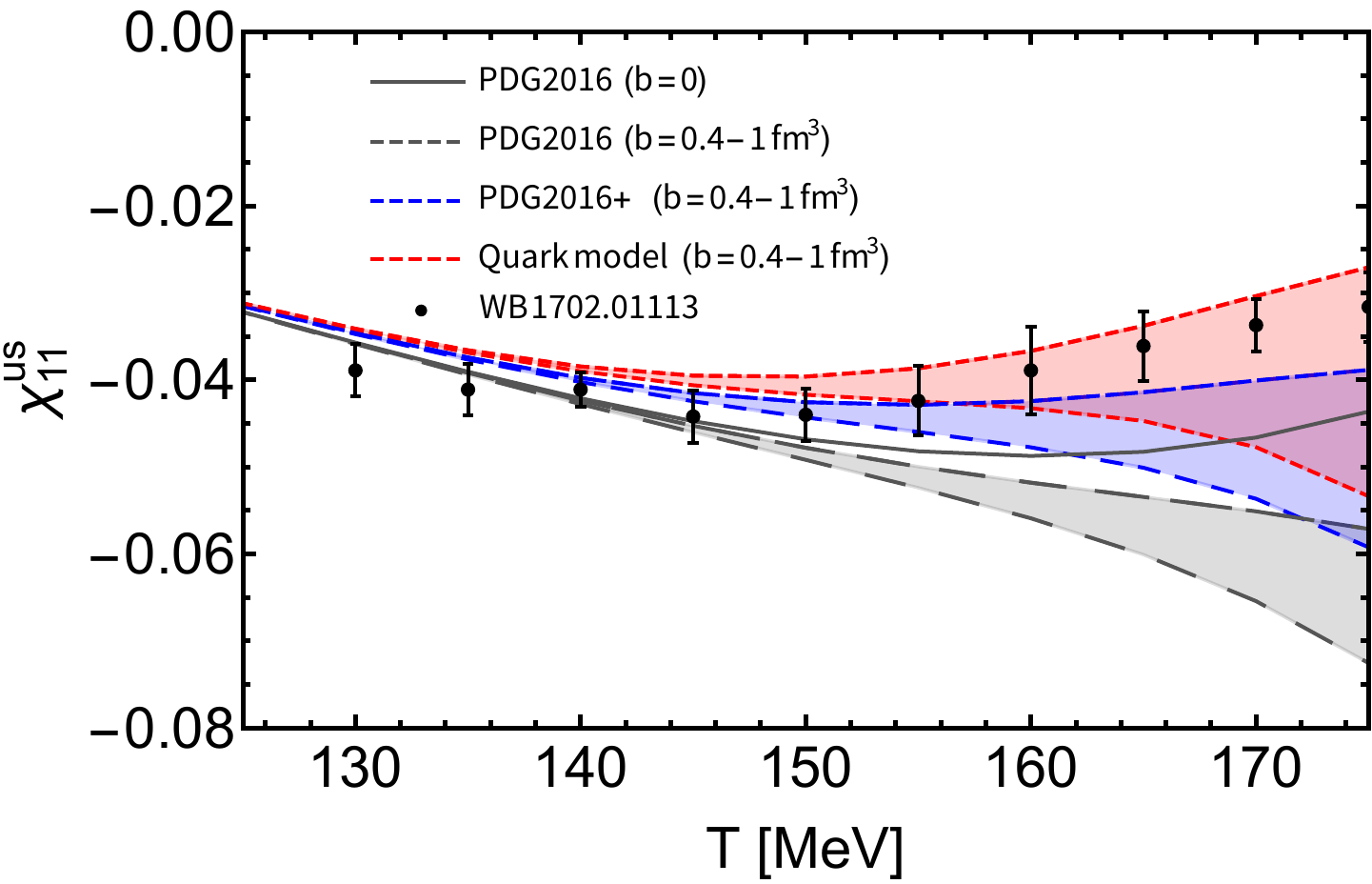}
    \caption{Up-strange quark susceptibility, $\chi^{us}_{11}$, as a function of the temperature. The continuum extrapolated lattice results are given as black points with error bars \cite{Alba:2017mqu}, while the EV-HRG calculations are shown for a range of the parameter $b$ between 0.4 and 1 $\text{fm}^3$ for different hadronic lists.}
    \label{fig:chi-us}
\end{figure}

Figure~\ref{fig:chi-4S-2S} depicts the temperature dependence of $\chi_4^S/\chi_2^S$.
This quantity does not involve any $\mu_B$ derivatives, thus it does not probe the repulsive baryonic interactions as directly as e.g. $\chi_4^B/\chi_2^B$ that we considered before.
Due to that fact, we calculate over a broader range of $b = 0-1$~fm$^3$ for the EV parameter value, given by the bands in Fig.~\ref{fig:chi-4S-2S}.
Since there are no $\mu_B$ derivatives, both hyperons as well as strange mesons like kaons contribute to this quantity.
Moreover, the values of $\chi_4^S/\chi_2^S$ above unity are due to multi-strange hyperons.
Figure~\ref{fig:chi-4S-2S} shows that the excluded volume suppresses $\chi_4^S/\chi_2^S$. This effect is less pronounced at smaller temperatures, but becomes sizable at $T \gtrsim 150$~MeV.
We see that the lattice data are underestimated when using the standard PDG2016 list, and the excluded volume does not improve the agreement.
For the PDG2016+ list the agreement with the lattice data is obtained at $T \lesssim 160$~MeV for the smallest considered values of the excluded volume~(the upper part of the blue band).
This is consistent with the observation made earlier that strange baryons prefer a smaller excluded volume.
On the other hand, when the QM list is considered, which contains even more extra strange baryons, the best agreement with the lattice data is obtained for the higher value of the EV parameter, $b \simeq 1$~fm$^3$~(the lower part of the red band).
We see similar effects for  $\chi^{us}_{11}$, shown in Fig.~\ref{fig:chi-us}, with a slight over-prediction in the case of the QM list.
The PDG2016 list describes the lattice data at $T \lesssim 150$~MeV, but breaks down at higher temperatures, with no benefit from introducing the excluded volume.
The PDG2016+ and QM lists allow us to extend the agreement with the lattice data for $\chi^{us}_{11}$ to $T = 160-165$~MeV when baryon excluded volumes of up to $b \simeq 0.4$~fm$^3$ and $b \simeq 1$~fm$^3$ are used, respectively.

\begin{figure*}
    \centering
    \includegraphics[width=0.48\linewidth]{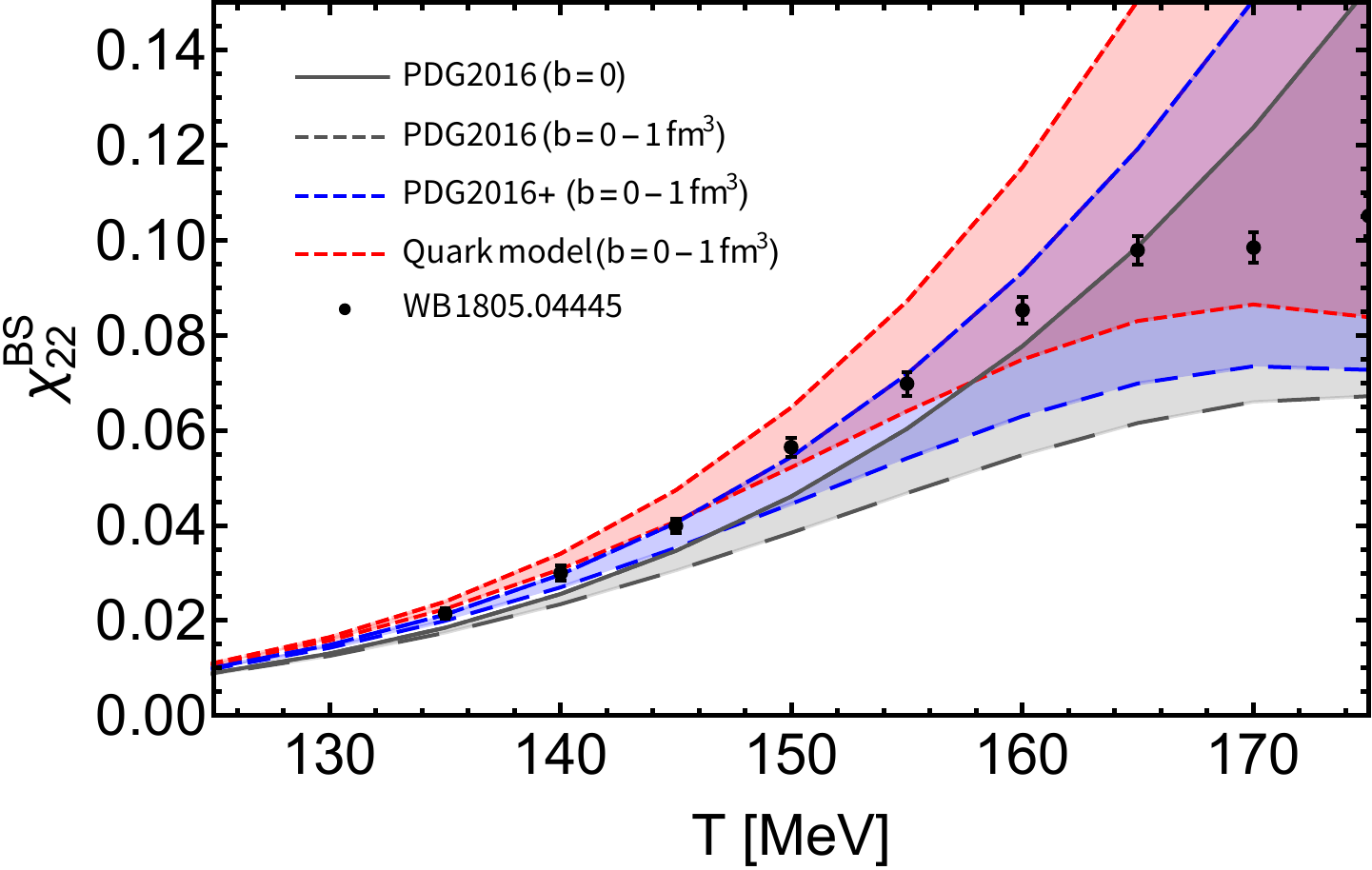}
    \includegraphics[width=0.48\linewidth]{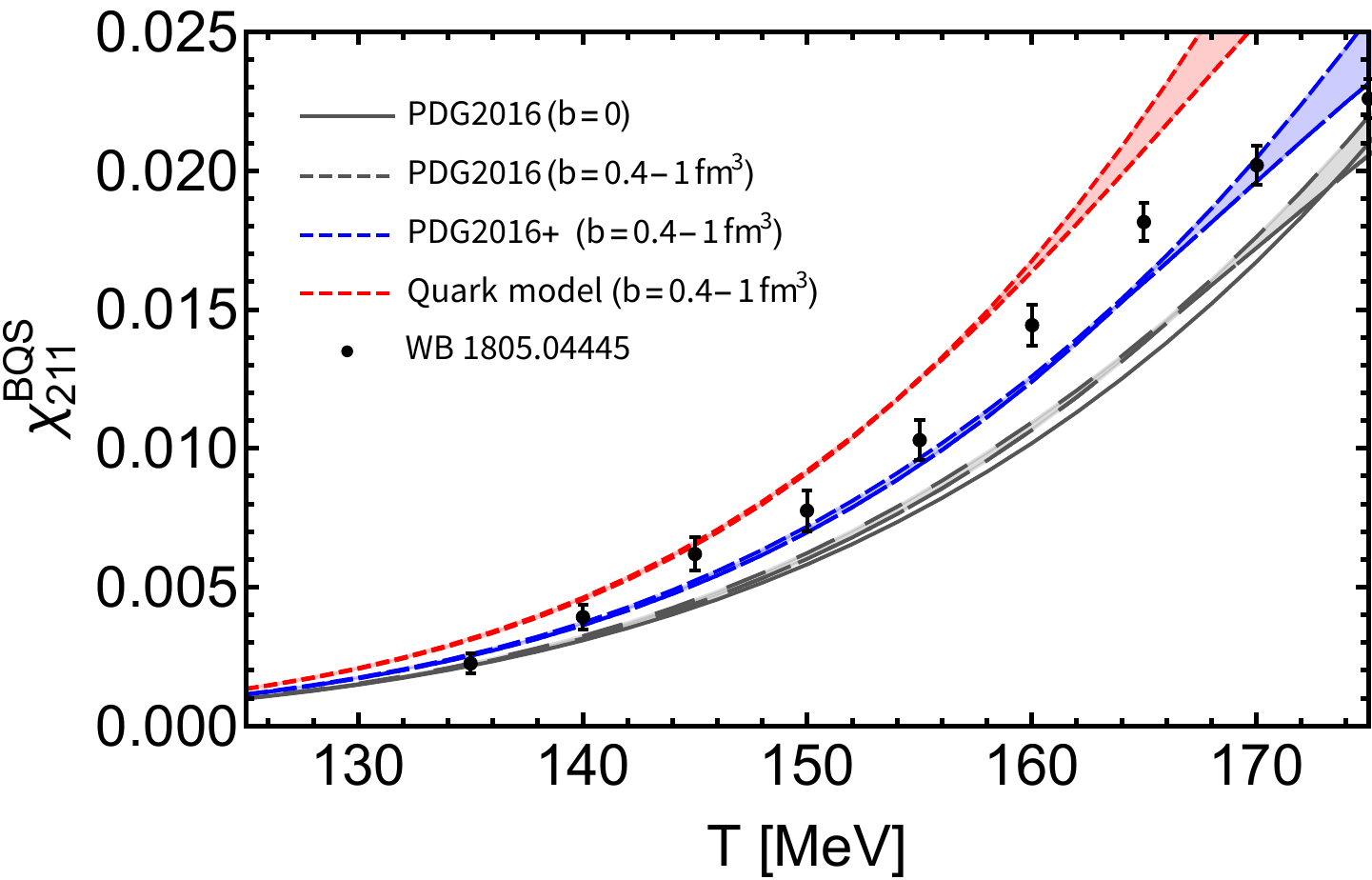}
    \caption{Temperature dependence of mixed fourth order susceptibilities $\chi^{BS}_{22}$ and $\chi^{BQS}_{211}$. The results at finite lattice spacing $N_{\tau}=12$ from Ref. \cite{Borsanyi:2018grb} are shown as black points with error bars. Left: Fourth-order baryon-strangeness susceptibility calculated with several different hadronic lists in the EV-HRG model for the full range of the EV parameter $b$. Right: Mixed fourth order $BQS$ susceptibility calculated with several different hadronic lists in the EV-HRG model for a range of the EV parameter $b$ between 0.4 and 1~fm$^3$.}
    \label{fig:chi-22BS}
\end{figure*}

Based on the analysis of $\chi_4^S/\chi_2^S$ and $\chi^{us}_{11}$, we observe a correlation between the number of extra strange states and baryon excluded volume: the larger the number of extra states, the larger the excluded volume must be in order to recover agreement with the lattice data for these two quantities.
Thus the attractive interactions via the inclusion of extra states can be balanced by an additional repulsion in the baryon sector in these observables.
One way to break this degeneracy is to consider quantities that probe only a single one of the effects, like the $\chi_{11}^{BQ}/\chi_2^B$ and $\chi_{11}^{BS}/\chi_2^B$ ratios that we considered earlier.
However, those ratios may be sensitive to additional physics like modeling of broad resonances, as discussed above.
Instead, we study here  a couple of additional susceptibilities that probe the strangeness content of baryons.
First we look at $\chi^{BS}_{22}$, which is a doubly strange quantity and thus more sensitive to multi-strange hyperons.
This quantity, shown in the left panel of Fig.~\ref{fig:chi-22BS}, paints a picture consistent with $\chi_4^S/\chi_2^S$ and $\chi^{us}_{11}$: extra states are clearly needed to describe the lattice data, but PDG2016+ prefers a smaller excluded volume while the QM list prefers a larger one.
On the other hand, the mixed $BQS$ susceptibility containing only one strangeness derivative $\chi^{BQS}_{211}$, shown in the right panel of Fig.~\ref{fig:chi-22BS}, exhibits only mild dependence on the excluded volume
but large sensitivity to the number of extra states. 
The minimal dependence on the excluded volume parameter $b$ of this fourth order susceptibility can be understood when one considers that the derivatives with respect to $Q$ and $S$ are only single derivatives. 
Some baryons, e.g. $\Sigma^+$, carry these conserved charges with opposite signs~(see Table~\ref{tab:hadron_table}).
For similar reasons, the contributions of the various baryon-baryon interactions to $\chi^{BQS}_{211}$ can be either positive or negative and as such, each term making up this quantity, as evaluated from Eqs. \eqref{eq:pBevW} and \eqref{eq:chis}, can contribute with different signs. 
This leads to a smaller $b$-dependence of $\chi^{BQS}_{211}$ than that of $\chi^{BS}_{22}$~(the left panel of Fig.~\ref{fig:chi-22BS}), in which case all contributions carry the same sign due to the fact that both the baryon number $B$ and the strangeness $S$ in this observable are squared.
The lattice results tend to lie in between the predictions based on the PDG2016+ and QM lists.
This indicates that the number of extra states might be underestimated in the PDG2016+ list but overestimated in the QM list.
Thus, the difference between the two lists may be taken as a systematic uncertainty in the particle list.

We note that the lattice data on
the susceptibilities $\chi^{BS}_{22}$ and $\chi^{BQS}_{211}$, as well as the ratios $\chi_{31}^{BS}/\chi_{11}^{BS}$ and $\chi_{31}^{BQ}/\chi_{11}^{BQ}$, are not continuum extrapolated but only available up to a lattice spacing of $N_{\tau} = 12$. While continuum extrapolated lattice results for $\chi_4^B/\chi_2^B$ have been shown e.g. in Ref. \cite{Borsanyi:2013hza}, in this manuscript we use $N_{\tau}=12$ also for this quantity, for a consistent comparison with the other ratios.
These data could be subject to slight alterations in the continuum limit, in particular at the lower temperatures.
Additionally, there are many more susceptibilities available from the lattice than those explored in this work. 
However, in this manuscript we focused mostly on baryon and strangeness observables in order to probe the extra strange states and baryon-baryon interactions. 
This allows us to avoid the most severe lattice systematics, for instance taste violation, which mainly affect calculations sensitive to pion degrees of freedom, like e.g. electric charge susceptibilities.

\section{Conclusions}\label{sec:concl}

We have investigated two common extensions of the Hadron Resonance Gas model 
that implement additional attractive and repulsive interactions among hadrons.
The attractive interactions correspond to adding extra
states exceeding those measured with high confidence by the Particle Data Group, leading to additive corrections to the overall pressure in the HRG model. 
This has been studied with the use of the PDG2016+ and QM particle lists. 
On the other hand, we also apply excluded-volume corrections in the baryon sector, 
which model the presence of repulsive core in (anti)baryon-(anti)baryon interactions. 
We demonstrate that these two extensions are complementary and find support in the available first-principles lattice QCD data.

To constrain the two extensions simultaneously we constructed specific combinations of conserved charge susceptibilities that probe the two effects separately.
We show that the second order ratios, $\chi_{11}^{BS}/\chi_2^B$ and $\chi_{11}^{BQ}/\chi_2^B$, probe only the hadron spectrum but not the excluded volume. 
The inclusion of additional states improves the agreement with the lattice data for these two quantities, with the best description obtained using the QM list.
Even for the QM list, however, deviations from the lattice data emerge at $T \sim 150-155$~MeV.
We argued that these deviations may necessitate a more involved modeling of broad resonances as well the possibility of smaller excluded volumes for strange baryons compared to non-strange ones.
We further studied the constraints on the hadronic spectrum by analyzing the various strangeness susceptibilities, including $\chi_4^S/\chi_2^S$, $\chi^{us}_{11}$, $\chi^{BS}_{22}$ and $\chi^{BQS}_{211}$.
The analysis of the lattice data for these quantities indicates that both the PDG2016+ and QM lists are preferable over the standard list, where the former contains most but not all the extra strange states while the QM list contains too many.
Therefore, the difference between the results using these two lists can be taken as a systematic uncertainty due to the hadron spectrum.

The fourth-to-second order ratios, $\chi_4^B/\chi_2^B$, $\chi_{31}^{BS}/\chi_{11}^{BS}$, and $\chi_{31}^{BQ}/\chi_{11}^{BQ}$ are shown to be mainly sensitive to the excluded volume corrections and, thus, suitable to constrain these corrections.
In the absence of the excluded volume corrections, these three ratios are equal to unity irrespective of the hadronic spectrum.
The excluded volume effects, on the other hand, suppress the ratios and make them behave similarly to the lattice data at $T \sim 155-165$~MeV.
The EV-HRG model that we use assigns a constant EV parameter $b$ for all baryons and predicts $\chi_4^B/\chi_2^B$, $\chi_{31}^{BS}/\chi_{11}^{BS}$, and $\chi_{31}^{BQ}/\chi_{11}^{BQ}$ to be all equal to one another at a given temperature.
The lattice data, on the other hand, reveal statistically significant differences between the three ratios at $T \gtrsim 155$~MeV, which follow a hierarchy $\chi_{31}^{BS}/\chi_{11}^{BS} > \chi_4^B/\chi_2^B > \chi_{31}^{BQ}/\chi_{11}^{BQ}$.
We argued that this hierarchy indicates a flavor dependence in the baryon excluded volumes, namely that strange baryons have generally smaller excluded volumes than non-strange baryons.
Therefore, while using a constant EV parameter in a range $b = 0.4-1$~fm$^3$ may be good enough to capture the general suppression of $\chi_4^B/\chi_2^B$, $\chi_{31}^{BS}/\chi_{11}^{BS}$, and $\chi_{31}^{BQ}/\chi_{11}^{BQ}$ ratios at $T \sim 155-165$~MeV, a more involved model is necessary to describe the subtle differences between the three.

In summary, an extended HRG model incorporating extra states via either the PDG2016+ or QM list as well as baryon excluded volume with parameter $b = 0.4-1$~fm$^3$ significantly improves the description of many lattice QCD susceptibilities at temperatures up to $T \simeq 160-165$~MeV over the standard HRG model. 
We note that the results from these two lists provide an estimate of the theoretical uncertainty on the hadronic list.
This is particularly relevant for the chemical freeze-out conditions realized in heavy-ion collisions and can be used for improved modeling of event-by-event fluctuations measured in the corresponding experiments at the LHC, RHIC, and SPS.
For instance, the EV-HRG model studied here can be directly used in the generalized Cooper-Frye particlization routine developed in Ref.~\cite{Vovchenko:2020kwg}.
We summarize the performance of each list as follows. We find that the QM list performs better for the ratio $\chi_{11}^{BS}/\chi_2^B$, while both lists describe the continuum extrapolated lattice data well for $\chi_{11}^{BQ}/\chi_2^B$ as shown in Fig. \ref{fig:chi-BS-2B_no_EV_dep}.
For ratios of fourth-to-second order susceptibilities as shown in Fig. \ref{fig:chi-4-2_equiv_EV}, we find, as we expect, a small variation in the results within the EV-HRG model for different lists.
From strangeness-sensitive susceptibilities $\chi_4^S/\chi_2^S$, $\chi_{11}^{us}$ and $\chi_{22}^{BS}$, we determined that the PDG2016+ list agrees best with the continuum extrapolated lattice data for a small excluded volume $b \simeq 0.2-0.4$~fm$^3$, while the QM list finds agreement when $b \simeq 1$~fm$^3$.
For the mixed fourth order susceptibility  $\chi_{211}^{BQS}$, the optimal list for treatment of the chemical freeze-out is PDG2016+.
Given that there is still some tension between the PDG2016+ list and the QM list, our results could indicate that more work is still needed to determine the number of states in the hadronic spectrum in heavy-ion collisions.
Further improvements of the model can be achieved by considering differences in excluded volumes of strange and non-strange baryons, as well as a more involved modeling of broad resonances.

\section*{Acknowledgements}
The authors would like to thank Paolo Parotto for discussions and help preparing the updated version of the QM list used in this publication.
This material is based upon work supported by the National Science Foundation under grant no. PHY1654219, by the U.S. Department of Energy, Office of Science, Office of Nuclear Physics, under contract number DE-AC02-05CH11231 and within the framework of the Beam Energy Scan Theory (BEST) Collaboration. 
V.V. acknowledges the support via the
Feodor Lynen program of the Alexander von Humboldt
foundation.

\bibliography{all}

\begin{thebibliography}{101}%
\makeatletter
\providecommand \@ifxundefined [1]{%
 \@ifx{#1\undefined}
}%
\providecommand \@ifnum [1]{%
 \ifnum #1\expandafter \@firstoftwo
 \else \expandafter \@secondoftwo
 \fi
}%
\providecommand \@ifx [1]{%
 \ifx #1\expandafter \@firstoftwo
 \else \expandafter \@secondoftwo
 \fi
}%
\providecommand \natexlab [1]{#1}%
\providecommand \enquote  [1]{``#1''}%
\providecommand \bibnamefont  [1]{#1}%
\providecommand \bibfnamefont [1]{#1}%
\providecommand \citenamefont [1]{#1}%
\providecommand \href@noop [0]{\@secondoftwo}%
\providecommand \href [0]{\begingroup \@sanitize@url \@href}%
\providecommand \@href[1]{\@@startlink{#1}\@@href}%
\providecommand \@@href[1]{\endgroup#1\@@endlink}%
\providecommand \@sanitize@url [0]{\catcode `\\12\catcode `\$12\catcode
  `\&12\catcode `\#12\catcode `\^12\catcode `\_12\catcode `\%12\relax}%
\providecommand \@@startlink[1]{}%
\providecommand \@@endlink[0]{}%
\providecommand \url  [0]{\begingroup\@sanitize@url \@url }%
\providecommand \@url [1]{\endgroup\@href {#1}{\urlprefix }}%
\providecommand \urlprefix  [0]{URL }%
\providecommand \Eprint [0]{\href }%
\providecommand \doibase [0]{http://dx.doi.org/}%
\providecommand \selectlanguage [0]{\@gobble}%
\providecommand \bibinfo  [0]{\@secondoftwo}%
\providecommand \bibfield  [0]{\@secondoftwo}%
\providecommand \translation [1]{[#1]}%
\providecommand \BibitemOpen [0]{}%
\providecommand \bibitemStop [0]{}%
\providecommand \bibitemNoStop [0]{.\EOS\space}%
\providecommand \EOS [0]{\spacefactor3000\relax}%
\providecommand \BibitemShut  [1]{\csname bibitem#1\endcsname}%
\let\auto@bib@innerbib\@empty
\bibitem [{\citenamefont {Aoki}\ \emph
  {et~al.}(2006{\natexlab{a}})\citenamefont {Aoki}, \citenamefont {Endrodi},
  \citenamefont {Fodor}, \citenamefont {Katz},\ and\ \citenamefont
  {Szabo}}]{Aoki:2006we}%
  \BibitemOpen
  \bibfield  {author} {\bibinfo {author} {\bibfnamefont {Y.}~\bibnamefont
  {Aoki}}, \bibinfo {author} {\bibfnamefont {G.}~\bibnamefont {Endrodi}},
  \bibinfo {author} {\bibfnamefont {Z.}~\bibnamefont {Fodor}}, \bibinfo
  {author} {\bibfnamefont {S.~D.}\ \bibnamefont {Katz}}, \ and\ \bibinfo
  {author} {\bibfnamefont {K.~K.}\ \bibnamefont {Szabo}},\ }\href {\doibase
  10.1038/nature05120} {\bibfield  {journal} {\bibinfo  {journal} {Nature}\
  }\textbf {\bibinfo {volume} {443}},\ \bibinfo {pages} {675} (\bibinfo {year}
  {2006}{\natexlab{a}})},\ \Eprint {http://arxiv.org/abs/hep-lat/0611014}
  {arXiv:hep-lat/0611014 [hep-lat]} \BibitemShut {NoStop}%
\bibitem [{\citenamefont {Aoki}\ \emph
  {et~al.}(2006{\natexlab{b}})\citenamefont {Aoki}, \citenamefont {Fodor},
  \citenamefont {Katz},\ and\ \citenamefont {Szabo}}]{Aoki:2006br}%
  \BibitemOpen
  \bibfield  {author} {\bibinfo {author} {\bibfnamefont {Y.}~\bibnamefont
  {Aoki}}, \bibinfo {author} {\bibfnamefont {Z.}~\bibnamefont {Fodor}},
  \bibinfo {author} {\bibfnamefont {S.~D.}\ \bibnamefont {Katz}}, \ and\
  \bibinfo {author} {\bibfnamefont {K.~K.}\ \bibnamefont {Szabo}},\ }\href
  {\doibase 10.1016/j.physletb.2006.10.021} {\bibfield  {journal} {\bibinfo
  {journal} {Phys. Lett.}\ }\textbf {\bibinfo {volume} {B643}},\ \bibinfo
  {pages} {46} (\bibinfo {year} {2006}{\natexlab{b}})},\ \Eprint
  {http://arxiv.org/abs/hep-lat/0609068} {arXiv:hep-lat/0609068 [hep-lat]}
  \BibitemShut {NoStop}%
\bibitem [{\citenamefont {Borsanyi}\ \emph
  {et~al.}(2010{\natexlab{a}})\citenamefont {Borsanyi}, \citenamefont {Fodor},
  \citenamefont {Hoelbling}, \citenamefont {Katz}, \citenamefont {Krieg},
  \citenamefont {Ratti},\ and\ \citenamefont {Szabo}}]{Borsanyi:2010bp}%
  \BibitemOpen
  \bibfield  {author} {\bibinfo {author} {\bibfnamefont {S.}~\bibnamefont
  {Borsanyi}}, \bibinfo {author} {\bibfnamefont {Z.}~\bibnamefont {Fodor}},
  \bibinfo {author} {\bibfnamefont {C.}~\bibnamefont {Hoelbling}}, \bibinfo
  {author} {\bibfnamefont {S.~D.}\ \bibnamefont {Katz}}, \bibinfo {author}
  {\bibfnamefont {S.}~\bibnamefont {Krieg}}, \bibinfo {author} {\bibfnamefont
  {C.}~\bibnamefont {Ratti}}, \ and\ \bibinfo {author} {\bibfnamefont {K.~K.}\
  \bibnamefont {Szabo}} (\bibinfo {collaboration} {Wuppertal-Budapest}),\
  }\href {\doibase 10.1007/JHEP09(2010)073} {\bibfield  {journal} {\bibinfo
  {journal} {JHEP}\ }\textbf {\bibinfo {volume} {09}},\ \bibinfo {pages} {073}
  (\bibinfo {year} {2010}{\natexlab{a}})},\ \Eprint
  {http://arxiv.org/abs/1005.3508} {arXiv:1005.3508 [hep-lat]} \BibitemShut
  {NoStop}%
\bibitem [{\citenamefont {Bazavov}\ \emph
  {et~al.}(2012{\natexlab{a}})\citenamefont {Bazavov} \emph
  {et~al.}}]{Bazavov:2011nk}%
  \BibitemOpen
  \bibfield  {author} {\bibinfo {author} {\bibfnamefont {A.}~\bibnamefont
  {Bazavov}} \emph {et~al.},\ }\href {\doibase 10.1103/PhysRevD.85.054503}
  {\bibfield  {journal} {\bibinfo  {journal} {Phys. Rev. D}\ }\textbf {\bibinfo
  {volume} {85}},\ \bibinfo {pages} {054503} (\bibinfo {year}
  {2012}{\natexlab{a}})},\ \Eprint {http://arxiv.org/abs/1111.1710}
  {arXiv:1111.1710 [hep-lat]} \BibitemShut {NoStop}%
\bibitem [{\citenamefont {Borsanyi}\ \emph
  {et~al.}(2014{\natexlab{a}})\citenamefont {Borsanyi}, \citenamefont {Fodor},
  \citenamefont {Hoelbling}, \citenamefont {Katz}, \citenamefont {Krieg},\ and\
  \citenamefont {Szabo}}]{Borsanyi:2013bia}%
  \BibitemOpen
  \bibfield  {author} {\bibinfo {author} {\bibfnamefont {S.}~\bibnamefont
  {Borsanyi}}, \bibinfo {author} {\bibfnamefont {Z.}~\bibnamefont {Fodor}},
  \bibinfo {author} {\bibfnamefont {C.}~\bibnamefont {Hoelbling}}, \bibinfo
  {author} {\bibfnamefont {S.~D.}\ \bibnamefont {Katz}}, \bibinfo {author}
  {\bibfnamefont {S.}~\bibnamefont {Krieg}}, \ and\ \bibinfo {author}
  {\bibfnamefont {K.~K.}\ \bibnamefont {Szabo}},\ }\href {\doibase
  10.1016/j.physletb.2014.01.007} {\bibfield  {journal} {\bibinfo  {journal}
  {Phys. Lett.}\ }\textbf {\bibinfo {volume} {B730}},\ \bibinfo {pages} {99}
  (\bibinfo {year} {2014}{\natexlab{a}})},\ \Eprint
  {http://arxiv.org/abs/1309.5258} {arXiv:1309.5258 [hep-lat]} \BibitemShut
  {NoStop}%
\bibitem [{\citenamefont {Bazavov}\ \emph
  {et~al.}(2014{\natexlab{a}})\citenamefont {Bazavov} \emph
  {et~al.}}]{Bazavov:2014pvz}%
  \BibitemOpen
  \bibfield  {author} {\bibinfo {author} {\bibfnamefont {A.}~\bibnamefont
  {Bazavov}} \emph {et~al.} (\bibinfo {collaboration} {HotQCD}),\ }\href
  {\doibase 10.1103/PhysRevD.90.094503} {\bibfield  {journal} {\bibinfo
  {journal} {Phys. Rev.}\ }\textbf {\bibinfo {volume} {D90}},\ \bibinfo {pages}
  {094503} (\bibinfo {year} {2014}{\natexlab{a}})},\ \Eprint
  {http://arxiv.org/abs/1407.6387} {arXiv:1407.6387 [hep-lat]} \BibitemShut
  {NoStop}%
\bibitem [{\citenamefont {Fodor}\ \emph {et~al.}(2019)\citenamefont {Fodor},
  \citenamefont {Giordano}, \citenamefont {G\"unther}, \citenamefont {Kap\'as},
  \citenamefont {Katz}, \citenamefont {P\'asztor}, \citenamefont {Portillo},
  \citenamefont {Ratti}, \citenamefont {Sexty},\ and\ \citenamefont
  {Szab\'o}}]{Fodor:2018wul}%
  \BibitemOpen
  \bibfield  {author} {\bibinfo {author} {\bibfnamefont {Z.}~\bibnamefont
  {Fodor}}, \bibinfo {author} {\bibfnamefont {M.}~\bibnamefont {Giordano}},
  \bibinfo {author} {\bibfnamefont {J.~N.}\ \bibnamefont {G\"unther}}, \bibinfo
  {author} {\bibfnamefont {K.}~\bibnamefont {Kap\'as}}, \bibinfo {author}
  {\bibfnamefont {S.~D.}\ \bibnamefont {Katz}}, \bibinfo {author}
  {\bibfnamefont {A.}~\bibnamefont {P\'asztor}}, \bibinfo {author}
  {\bibfnamefont {I.}~\bibnamefont {Portillo}}, \bibinfo {author}
  {\bibfnamefont {C.}~\bibnamefont {Ratti}}, \bibinfo {author} {\bibfnamefont
  {D.}~\bibnamefont {Sexty}}, \ and\ \bibinfo {author} {\bibfnamefont {K.~K.}\
  \bibnamefont {Szab\'o}},\ }\href {\doibase 10.1016/j.nuclphysa.2018.12.015}
  {\bibfield  {journal} {\bibinfo  {journal} {Nucl. Phys. A}\ }\textbf
  {\bibinfo {volume} {982}},\ \bibinfo {pages} {843} (\bibinfo {year}
  {2019})},\ \Eprint {http://arxiv.org/abs/1807.09862} {arXiv:1807.09862
  [hep-lat]} \BibitemShut {NoStop}%
\bibitem [{\citenamefont {Busza}\ \emph {et~al.}(2018)\citenamefont {Busza},
  \citenamefont {Rajagopal},\ and\ \citenamefont {van~der
  Schee}}]{Busza:2018rrf}%
  \BibitemOpen
  \bibfield  {author} {\bibinfo {author} {\bibfnamefont {W.}~\bibnamefont
  {Busza}}, \bibinfo {author} {\bibfnamefont {K.}~\bibnamefont {Rajagopal}}, \
  and\ \bibinfo {author} {\bibfnamefont {W.}~\bibnamefont {van~der Schee}},\
  }\href {\doibase 10.1146/annurev-nucl-101917-020852} {\bibfield  {journal}
  {\bibinfo  {journal} {Ann. Rev. Nucl. Part. Sci.}\ }\textbf {\bibinfo
  {volume} {68}},\ \bibinfo {pages} {339} (\bibinfo {year} {2018})},\ \Eprint
  {http://arxiv.org/abs/1802.04801} {arXiv:1802.04801 [hep-ph]} \BibitemShut
  {NoStop}%
\bibitem [{\citenamefont {Stephanov}(2011)}]{Stephanov:2011pb}%
  \BibitemOpen
  \bibfield  {author} {\bibinfo {author} {\bibfnamefont {M.}~\bibnamefont
  {Stephanov}},\ }\href {\doibase 10.1103/PhysRevLett.107.052301} {\bibfield
  {journal} {\bibinfo  {journal} {Phys. Rev. Lett.}\ }\textbf {\bibinfo
  {volume} {107}},\ \bibinfo {pages} {052301} (\bibinfo {year} {2011})},\
  \Eprint {http://arxiv.org/abs/1104.1627} {arXiv:1104.1627 [hep-ph]}
  \BibitemShut {NoStop}%
\bibitem [{\citenamefont {Stephanov}\ \emph {et~al.}(1998)\citenamefont
  {Stephanov}, \citenamefont {Rajagopal},\ and\ \citenamefont
  {Shuryak}}]{Stephanov:1998dy}%
  \BibitemOpen
  \bibfield  {author} {\bibinfo {author} {\bibfnamefont {M.~A.}\ \bibnamefont
  {Stephanov}}, \bibinfo {author} {\bibfnamefont {K.}~\bibnamefont
  {Rajagopal}}, \ and\ \bibinfo {author} {\bibfnamefont {E.~V.}\ \bibnamefont
  {Shuryak}},\ }\href {\doibase 10.1103/PhysRevLett.81.4816} {\bibfield
  {journal} {\bibinfo  {journal} {Phys. Rev. Lett.}\ }\textbf {\bibinfo
  {volume} {81}},\ \bibinfo {pages} {4816} (\bibinfo {year} {1998})},\ \Eprint
  {http://arxiv.org/abs/hep-ph/9806219} {arXiv:hep-ph/9806219} \BibitemShut
  {NoStop}%
\bibitem [{\citenamefont {Ratti}\ \emph {et~al.}(2006)\citenamefont {Ratti},
  \citenamefont {Thaler},\ and\ \citenamefont {Weise}}]{Ratti:2006gh}%
  \BibitemOpen
  \bibfield  {author} {\bibinfo {author} {\bibfnamefont {C.}~\bibnamefont
  {Ratti}}, \bibinfo {author} {\bibfnamefont {M.~A.}\ \bibnamefont {Thaler}}, \
  and\ \bibinfo {author} {\bibfnamefont {W.}~\bibnamefont {Weise}},\
  }\href@noop {} {\  (\bibinfo {year} {2006})},\ \Eprint
  {http://arxiv.org/abs/nucl-th/0604025} {arXiv:nucl-th/0604025} \BibitemShut
  {NoStop}%
\bibitem [{\citenamefont {Fukushima}\ and\ \citenamefont
  {Hatsuda}(2011)}]{Fukushima:2010bq}%
  \BibitemOpen
  \bibfield  {author} {\bibinfo {author} {\bibfnamefont {K.}~\bibnamefont
  {Fukushima}}\ and\ \bibinfo {author} {\bibfnamefont {T.}~\bibnamefont
  {Hatsuda}},\ }\href {\doibase 10.1088/0034-4885/74/1/014001} {\bibfield
  {journal} {\bibinfo  {journal} {Rept. Prog. Phys.}\ }\textbf {\bibinfo
  {volume} {74}},\ \bibinfo {pages} {014001} (\bibinfo {year} {2011})},\
  \Eprint {http://arxiv.org/abs/1005.4814} {arXiv:1005.4814 [hep-ph]}
  \BibitemShut {NoStop}%
\bibitem [{\citenamefont {Critelli}\ \emph {et~al.}(2017)\citenamefont
  {Critelli}, \citenamefont {Noronha}, \citenamefont {Noronha-Hostler},
  \citenamefont {Portillo}, \citenamefont {Ratti},\ and\ \citenamefont
  {Rougemont}}]{Critelli:2017oub}%
  \BibitemOpen
  \bibfield  {author} {\bibinfo {author} {\bibfnamefont {R.}~\bibnamefont
  {Critelli}}, \bibinfo {author} {\bibfnamefont {J.}~\bibnamefont {Noronha}},
  \bibinfo {author} {\bibfnamefont {J.}~\bibnamefont {Noronha-Hostler}},
  \bibinfo {author} {\bibfnamefont {I.}~\bibnamefont {Portillo}}, \bibinfo
  {author} {\bibfnamefont {C.}~\bibnamefont {Ratti}}, \ and\ \bibinfo {author}
  {\bibfnamefont {R.}~\bibnamefont {Rougemont}},\ }\href {\doibase
  10.1103/PhysRevD.96.096026} {\bibfield  {journal} {\bibinfo  {journal} {Phys.
  Rev. D}\ }\textbf {\bibinfo {volume} {96}},\ \bibinfo {pages} {096026}
  (\bibinfo {year} {2017})},\ \Eprint {http://arxiv.org/abs/1706.00455}
  {arXiv:1706.00455 [nucl-th]} \BibitemShut {NoStop}%
\bibitem [{\citenamefont {Luo}\ and\ \citenamefont {Xu}(2017)}]{Luo:2017faz}%
  \BibitemOpen
  \bibfield  {author} {\bibinfo {author} {\bibfnamefont {X.}~\bibnamefont
  {Luo}}\ and\ \bibinfo {author} {\bibfnamefont {N.}~\bibnamefont {Xu}},\
  }\href {\doibase 10.1007/s41365-017-0257-0} {\bibfield  {journal} {\bibinfo
  {journal} {Nucl. Sci. Tech.}\ }\textbf {\bibinfo {volume} {28}},\ \bibinfo
  {pages} {112} (\bibinfo {year} {2017})},\ \Eprint
  {http://arxiv.org/abs/1701.02105} {arXiv:1701.02105 [nucl-ex]} \BibitemShut
  {NoStop}%
\bibitem [{\citenamefont {Parotto}\ \emph {et~al.}(2020)\citenamefont
  {Parotto}, \citenamefont {Bluhm}, \citenamefont {Mroczek}, \citenamefont
  {Nahrgang}, \citenamefont {Noronha-Hostler}, \citenamefont {Rajagopal},
  \citenamefont {Ratti}, \citenamefont {Sch\"afer},\ and\ \citenamefont
  {Stephanov}}]{Parotto:2018pwx}%
  \BibitemOpen
  \bibfield  {author} {\bibinfo {author} {\bibfnamefont {P.}~\bibnamefont
  {Parotto}}, \bibinfo {author} {\bibfnamefont {M.}~\bibnamefont {Bluhm}},
  \bibinfo {author} {\bibfnamefont {D.}~\bibnamefont {Mroczek}}, \bibinfo
  {author} {\bibfnamefont {M.}~\bibnamefont {Nahrgang}}, \bibinfo {author}
  {\bibfnamefont {J.}~\bibnamefont {Noronha-Hostler}}, \bibinfo {author}
  {\bibfnamefont {K.}~\bibnamefont {Rajagopal}}, \bibinfo {author}
  {\bibfnamefont {C.}~\bibnamefont {Ratti}}, \bibinfo {author} {\bibfnamefont
  {T.}~\bibnamefont {Sch\"afer}}, \ and\ \bibinfo {author} {\bibfnamefont
  {M.}~\bibnamefont {Stephanov}},\ }\href {\doibase
  10.1103/PhysRevC.101.034901} {\bibfield  {journal} {\bibinfo  {journal}
  {Phys. Rev. C}\ }\textbf {\bibinfo {volume} {101}},\ \bibinfo {pages}
  {034901} (\bibinfo {year} {2020})},\ \Eprint
  {http://arxiv.org/abs/1805.05249} {arXiv:1805.05249 [hep-ph]} \BibitemShut
  {NoStop}%
\bibitem [{\citenamefont {Grefa}\ \emph {et~al.}(2021)\citenamefont {Grefa},
  \citenamefont {Noronha}, \citenamefont {Noronha-Hostler}, \citenamefont
  {Portillo}, \citenamefont {Ratti},\ and\ \citenamefont
  {Rougemont}}]{Grefa:2021qvt}%
  \BibitemOpen
  \bibfield  {author} {\bibinfo {author} {\bibfnamefont {J.}~\bibnamefont
  {Grefa}}, \bibinfo {author} {\bibfnamefont {J.}~\bibnamefont {Noronha}},
  \bibinfo {author} {\bibfnamefont {J.}~\bibnamefont {Noronha-Hostler}},
  \bibinfo {author} {\bibfnamefont {I.}~\bibnamefont {Portillo}}, \bibinfo
  {author} {\bibfnamefont {C.}~\bibnamefont {Ratti}}, \ and\ \bibinfo {author}
  {\bibfnamefont {R.}~\bibnamefont {Rougemont}},\ }\href@noop {} {\  (\bibinfo
  {year} {2021})},\ \Eprint {http://arxiv.org/abs/2102.12042} {arXiv:2102.12042
  [nucl-th]} \BibitemShut {NoStop}%
\bibitem [{\citenamefont {Karthein}\ \emph {et~al.}(2021)\citenamefont
  {Karthein}, \citenamefont {Mroczek}, \citenamefont {Nava~Acuna},
  \citenamefont {Noronha-Hostler}, \citenamefont {Parotto}, \citenamefont
  {Price},\ and\ \citenamefont {Ratti}}]{Karthein:2021nxe}%
  \BibitemOpen
  \bibfield  {author} {\bibinfo {author} {\bibfnamefont {J.~M.}\ \bibnamefont
  {Karthein}}, \bibinfo {author} {\bibfnamefont {D.}~\bibnamefont {Mroczek}},
  \bibinfo {author} {\bibfnamefont {A.~R.}\ \bibnamefont {Nava~Acuna}},
  \bibinfo {author} {\bibfnamefont {J.}~\bibnamefont {Noronha-Hostler}},
  \bibinfo {author} {\bibfnamefont {P.}~\bibnamefont {Parotto}}, \bibinfo
  {author} {\bibfnamefont {D.~R.~P.}\ \bibnamefont {Price}}, \ and\ \bibinfo
  {author} {\bibfnamefont {C.}~\bibnamefont {Ratti}},\ }\href {\doibase
  10.1140/epjp/s13360-021-01615-5} {\bibfield  {journal} {\bibinfo  {journal}
  {Eur. Phys. J. Plus}\ }\textbf {\bibinfo {volume} {136}},\ \bibinfo {pages}
  {621} (\bibinfo {year} {2021})},\ \Eprint {http://arxiv.org/abs/2103.08146}
  {arXiv:2103.08146 [hep-ph]} \BibitemShut {NoStop}%
\bibitem [{\citenamefont {Adamczyk}\ \emph {et~al.}(2017)\citenamefont
  {Adamczyk} \emph {et~al.}}]{Adamczyk:2017iwn}%
  \BibitemOpen
  \bibfield  {author} {\bibinfo {author} {\bibfnamefont {L.}~\bibnamefont
  {Adamczyk}} \emph {et~al.} (\bibinfo {collaboration} {STAR}),\ }\href
  {\doibase 10.1103/PhysRevC.96.044904} {\bibfield  {journal} {\bibinfo
  {journal} {Phys. Rev.}\ }\textbf {\bibinfo {volume} {C96}},\ \bibinfo {pages}
  {044904} (\bibinfo {year} {2017})},\ \Eprint
  {http://arxiv.org/abs/1701.07065} {arXiv:1701.07065 [nucl-ex]} \BibitemShut
  {NoStop}%
\bibitem [{\citenamefont {Adamczewski-Musch}\ \emph {et~al.}(2020)\citenamefont
  {Adamczewski-Musch} \emph {et~al.}}]{Adamczewski-Musch:2020slf}%
  \BibitemOpen
  \bibfield  {author} {\bibinfo {author} {\bibfnamefont {J.}~\bibnamefont
  {Adamczewski-Musch}} \emph {et~al.} (\bibinfo {collaboration} {HADES}),\
  }\href {\doibase 10.1103/PhysRevC.102.024914} {\bibfield  {journal} {\bibinfo
   {journal} {Phys. Rev. C}\ }\textbf {\bibinfo {volume} {102}},\ \bibinfo
  {pages} {024914} (\bibinfo {year} {2020})},\ \Eprint
  {http://arxiv.org/abs/2002.08701} {arXiv:2002.08701 [nucl-ex]} \BibitemShut
  {NoStop}%
\bibitem [{\citenamefont {Abelev}\ \emph {et~al.}(2013)\citenamefont {Abelev}
  \emph {et~al.}}]{Abelev:2013vea}%
  \BibitemOpen
  \bibfield  {author} {\bibinfo {author} {\bibfnamefont {B.}~\bibnamefont
  {Abelev}} \emph {et~al.} (\bibinfo {collaboration} {ALICE}),\ }\href
  {\doibase 10.1103/PhysRevC.88.044910} {\bibfield  {journal} {\bibinfo
  {journal} {Phys. Rev.}\ }\textbf {\bibinfo {volume} {C88}},\ \bibinfo {pages}
  {044910} (\bibinfo {year} {2013})},\ \Eprint {http://arxiv.org/abs/1303.0737}
  {arXiv:1303.0737 [hep-ex]} \BibitemShut {NoStop}%
\bibitem [{\citenamefont {Fu}\ \emph {et~al.}(2021)\citenamefont {Fu},
  \citenamefont {Luo}, \citenamefont {Pawlowski}, \citenamefont {Rennecke},
  \citenamefont {Wen},\ and\ \citenamefont {Yin}}]{Fu:2021oaw}%
  \BibitemOpen
  \bibfield  {author} {\bibinfo {author} {\bibfnamefont {W.-j.}\ \bibnamefont
  {Fu}}, \bibinfo {author} {\bibfnamefont {X.}~\bibnamefont {Luo}}, \bibinfo
  {author} {\bibfnamefont {J.~M.}\ \bibnamefont {Pawlowski}}, \bibinfo {author}
  {\bibfnamefont {F.}~\bibnamefont {Rennecke}}, \bibinfo {author}
  {\bibfnamefont {R.}~\bibnamefont {Wen}}, \ and\ \bibinfo {author}
  {\bibfnamefont {S.}~\bibnamefont {Yin}},\ }\href@noop {} {\  (\bibinfo {year}
  {2021})},\ \Eprint {http://arxiv.org/abs/2101.06035} {arXiv:2101.06035
  [hep-ph]} \BibitemShut {NoStop}%
\bibitem [{\citenamefont {Ding}\ \emph {et~al.}(2015)\citenamefont {Ding},
  \citenamefont {Karsch},\ and\ \citenamefont {Mukherjee}}]{Ding:2015ona}%
  \BibitemOpen
  \bibfield  {author} {\bibinfo {author} {\bibfnamefont {H.-T.}\ \bibnamefont
  {Ding}}, \bibinfo {author} {\bibfnamefont {F.}~\bibnamefont {Karsch}}, \ and\
  \bibinfo {author} {\bibfnamefont {S.}~\bibnamefont {Mukherjee}},\ }\href
  {\doibase 10.1142/S0218301315300076} {\bibfield  {journal} {\bibinfo
  {journal} {Int. J. Mod. Phys. E}\ }\textbf {\bibinfo {volume} {24}},\
  \bibinfo {pages} {1530007} (\bibinfo {year} {2015})},\ \Eprint
  {http://arxiv.org/abs/1504.05274} {arXiv:1504.05274 [hep-lat]} \BibitemShut
  {NoStop}%
\bibitem [{\citenamefont {Ratti}(2018)}]{Ratti:2018ksb}%
  \BibitemOpen
  \bibfield  {author} {\bibinfo {author} {\bibfnamefont {C.}~\bibnamefont
  {Ratti}},\ }\href {\doibase 10.1088/1361-6633/aabb97} {\bibfield  {journal}
  {\bibinfo  {journal} {Rept. Prog. Phys.}\ }\textbf {\bibinfo {volume} {81}},\
  \bibinfo {pages} {084301} (\bibinfo {year} {2018})},\ \Eprint
  {http://arxiv.org/abs/1804.07810} {arXiv:1804.07810 [hep-lat]} \BibitemShut
  {NoStop}%
\bibitem [{\citenamefont {Bzdak}\ \emph {et~al.}(2020)\citenamefont {Bzdak},
  \citenamefont {Esumi}, \citenamefont {Koch}, \citenamefont {Liao},
  \citenamefont {Stephanov},\ and\ \citenamefont {Xu}}]{Bzdak:2019pkr}%
  \BibitemOpen
  \bibfield  {author} {\bibinfo {author} {\bibfnamefont {A.}~\bibnamefont
  {Bzdak}}, \bibinfo {author} {\bibfnamefont {S.}~\bibnamefont {Esumi}},
  \bibinfo {author} {\bibfnamefont {V.}~\bibnamefont {Koch}}, \bibinfo {author}
  {\bibfnamefont {J.}~\bibnamefont {Liao}}, \bibinfo {author} {\bibfnamefont
  {M.}~\bibnamefont {Stephanov}}, \ and\ \bibinfo {author} {\bibfnamefont
  {N.}~\bibnamefont {Xu}},\ }\href {\doibase 10.1016/j.physrep.2020.01.005}
  {\bibfield  {journal} {\bibinfo  {journal} {Phys. Rept.}\ }\textbf {\bibinfo
  {volume} {853}},\ \bibinfo {pages} {1} (\bibinfo {year} {2020})},\ \Eprint
  {http://arxiv.org/abs/1906.00936} {arXiv:1906.00936 [nucl-th]} \BibitemShut
  {NoStop}%
\bibitem [{\citenamefont {Ratti}\ and\ \citenamefont
  {Bellwied}(2021)}]{Ratti:2021ubw}%
  \BibitemOpen
  \bibfield  {author} {\bibinfo {author} {\bibfnamefont {C.}~\bibnamefont
  {Ratti}}\ and\ \bibinfo {author} {\bibfnamefont {R.}~\bibnamefont
  {Bellwied}},\ }\href {\doibase 10.1007/978-3-030-67235-5} {\emph {\bibinfo
  {title} {{The Deconfinement Transition of QCD: Theory Meets Experiment}}}},\
  \bibinfo {series} {Lecture Notes in Physics}, Vol.\ \bibinfo {volume} {981}\
  (\bibinfo {year} {2021})\BibitemShut {NoStop}%
\bibitem [{\citenamefont {Borsanyi}\ \emph
  {et~al.}(2010{\natexlab{b}})\citenamefont {Borsanyi}, \citenamefont
  {Endrodi}, \citenamefont {Fodor}, \citenamefont {Jakovac}, \citenamefont
  {Katz}, \citenamefont {Krieg}, \citenamefont {Ratti},\ and\ \citenamefont
  {Szabo}}]{Borsanyi:2010cj}%
  \BibitemOpen
  \bibfield  {author} {\bibinfo {author} {\bibfnamefont {S.}~\bibnamefont
  {Borsanyi}}, \bibinfo {author} {\bibfnamefont {G.}~\bibnamefont {Endrodi}},
  \bibinfo {author} {\bibfnamefont {Z.}~\bibnamefont {Fodor}}, \bibinfo
  {author} {\bibfnamefont {A.}~\bibnamefont {Jakovac}}, \bibinfo {author}
  {\bibfnamefont {S.~D.}\ \bibnamefont {Katz}}, \bibinfo {author}
  {\bibfnamefont {S.}~\bibnamefont {Krieg}}, \bibinfo {author} {\bibfnamefont
  {C.}~\bibnamefont {Ratti}}, \ and\ \bibinfo {author} {\bibfnamefont {K.~K.}\
  \bibnamefont {Szabo}},\ }\href {\doibase 10.1007/JHEP11(2010)077} {\bibfield
  {journal} {\bibinfo  {journal} {JHEP}\ }\textbf {\bibinfo {volume} {11}},\
  \bibinfo {pages} {077} (\bibinfo {year} {2010}{\natexlab{b}})},\ \Eprint
  {http://arxiv.org/abs/1007.2580} {arXiv:1007.2580 [hep-lat]} \BibitemShut
  {NoStop}%
\bibitem [{\citenamefont {Guenther}\ \emph {et~al.}(2017)\citenamefont
  {Guenther}, \citenamefont {Bellwied}, \citenamefont {Borsanyi}, \citenamefont
  {Fodor}, \citenamefont {Katz}, \citenamefont {Pasztor}, \citenamefont
  {Ratti},\ and\ \citenamefont {Szab\'o}}]{Guenther:2017hnx}%
  \BibitemOpen
  \bibfield  {author} {\bibinfo {author} {\bibfnamefont {J.~N.}\ \bibnamefont
  {Guenther}}, \bibinfo {author} {\bibfnamefont {R.}~\bibnamefont {Bellwied}},
  \bibinfo {author} {\bibfnamefont {S.}~\bibnamefont {Borsanyi}}, \bibinfo
  {author} {\bibfnamefont {Z.}~\bibnamefont {Fodor}}, \bibinfo {author}
  {\bibfnamefont {S.~D.}\ \bibnamefont {Katz}}, \bibinfo {author}
  {\bibfnamefont {A.}~\bibnamefont {Pasztor}}, \bibinfo {author} {\bibfnamefont
  {C.}~\bibnamefont {Ratti}}, \ and\ \bibinfo {author} {\bibfnamefont {K.~K.}\
  \bibnamefont {Szab\'o}},\ }\href {\doibase 10.1016/j.nuclphysa.2017.05.044}
  {\bibfield  {journal} {\bibinfo  {journal} {Nucl. Phys. A}\ }\textbf
  {\bibinfo {volume} {967}},\ \bibinfo {pages} {720} (\bibinfo {year}
  {2017})},\ \Eprint {http://arxiv.org/abs/1607.02493} {arXiv:1607.02493
  [hep-lat]} \BibitemShut {NoStop}%
\bibitem [{\citenamefont {G\"unther}\ \emph {et~al.}(2017)\citenamefont
  {G\"unther}, \citenamefont {Bellwied}, \citenamefont {Borsanyi},
  \citenamefont {Fodor}, \citenamefont {Katz}, \citenamefont {Pasztor},\ and\
  \citenamefont {Ratti}}]{Gunther:2017sxn}%
  \BibitemOpen
  \bibfield  {author} {\bibinfo {author} {\bibfnamefont {J.}~\bibnamefont
  {G\"unther}}, \bibinfo {author} {\bibfnamefont {R.}~\bibnamefont {Bellwied}},
  \bibinfo {author} {\bibfnamefont {S.}~\bibnamefont {Borsanyi}}, \bibinfo
  {author} {\bibfnamefont {Z.}~\bibnamefont {Fodor}}, \bibinfo {author}
  {\bibfnamefont {S.~D.}\ \bibnamefont {Katz}}, \bibinfo {author}
  {\bibfnamefont {A.}~\bibnamefont {Pasztor}}, \ and\ \bibinfo {author}
  {\bibfnamefont {C.}~\bibnamefont {Ratti}},\ }\href {\doibase
  10.1051/epjconf/201713707008} {\bibfield  {journal} {\bibinfo  {journal} {EPJ
  Web Conf.}\ }\textbf {\bibinfo {volume} {137}},\ \bibinfo {pages} {07008}
  (\bibinfo {year} {2017})}\BibitemShut {NoStop}%
\bibitem [{\citenamefont {Bazavov}\ \emph
  {et~al.}(2017{\natexlab{a}})\citenamefont {Bazavov} \emph
  {et~al.}}]{Bazavov:2017dus}%
  \BibitemOpen
  \bibfield  {author} {\bibinfo {author} {\bibfnamefont {A.}~\bibnamefont
  {Bazavov}} \emph {et~al.},\ }\href {\doibase 10.1103/PhysRevD.95.054504}
  {\bibfield  {journal} {\bibinfo  {journal} {Phys. Rev. D}\ }\textbf {\bibinfo
  {volume} {95}},\ \bibinfo {pages} {054504} (\bibinfo {year}
  {2017}{\natexlab{a}})},\ \Eprint {http://arxiv.org/abs/1701.04325}
  {arXiv:1701.04325 [hep-lat]} \BibitemShut {NoStop}%
\bibitem [{\citenamefont {Bors\'anyi}\ \emph {et~al.}(2021)\citenamefont
  {Bors\'anyi}, \citenamefont {Fodor}, \citenamefont {Guenther}, \citenamefont
  {Kara}, \citenamefont {Katz}, \citenamefont {Parotto}, \citenamefont
  {P\'asztor}, \citenamefont {Ratti},\ and\ \citenamefont
  {Szab\'o}}]{Borsanyi:2021sxv}%
  \BibitemOpen
  \bibfield  {author} {\bibinfo {author} {\bibfnamefont {S.}~\bibnamefont
  {Bors\'anyi}}, \bibinfo {author} {\bibfnamefont {Z.}~\bibnamefont {Fodor}},
  \bibinfo {author} {\bibfnamefont {J.~N.}\ \bibnamefont {Guenther}}, \bibinfo
  {author} {\bibfnamefont {R.}~\bibnamefont {Kara}}, \bibinfo {author}
  {\bibfnamefont {S.~D.}\ \bibnamefont {Katz}}, \bibinfo {author}
  {\bibfnamefont {P.}~\bibnamefont {Parotto}}, \bibinfo {author} {\bibfnamefont
  {A.}~\bibnamefont {P\'asztor}}, \bibinfo {author} {\bibfnamefont
  {C.}~\bibnamefont {Ratti}}, \ and\ \bibinfo {author} {\bibfnamefont {K.~K.}\
  \bibnamefont {Szab\'o}},\ }\href {\doibase 10.1103/PhysRevLett.126.232001}
  {\bibfield  {journal} {\bibinfo  {journal} {Phys. Rev. Lett.}\ }\textbf
  {\bibinfo {volume} {126}},\ \bibinfo {pages} {232001} (\bibinfo {year}
  {2021})},\ \Eprint {http://arxiv.org/abs/2102.06660} {arXiv:2102.06660
  [hep-lat]} \BibitemShut {NoStop}%
\bibitem [{\citenamefont {Mondal}\ \emph {et~al.}(2021)\citenamefont {Mondal},
  \citenamefont {Mukherjee},\ and\ \citenamefont {Hegde}}]{Mondal:2021jxk}%
  \BibitemOpen
  \bibfield  {author} {\bibinfo {author} {\bibfnamefont {S.}~\bibnamefont
  {Mondal}}, \bibinfo {author} {\bibfnamefont {S.}~\bibnamefont {Mukherjee}}, \
  and\ \bibinfo {author} {\bibfnamefont {P.}~\bibnamefont {Hegde}},\
  }\href@noop {} {\  (\bibinfo {year} {2021})},\ \Eprint
  {http://arxiv.org/abs/2106.03165} {arXiv:2106.03165 [hep-lat]} \BibitemShut
  {NoStop}%
\bibitem [{\citenamefont {Bellwied}\ \emph
  {et~al.}(2015{\natexlab{a}})\citenamefont {Bellwied}, \citenamefont
  {Borsanyi}, \citenamefont {Fodor}, \citenamefont {G\"unther}, \citenamefont
  {Katz}, \citenamefont {Ratti},\ and\ \citenamefont
  {Szabo}}]{Bellwied:2015rza}%
  \BibitemOpen
  \bibfield  {author} {\bibinfo {author} {\bibfnamefont {R.}~\bibnamefont
  {Bellwied}}, \bibinfo {author} {\bibfnamefont {S.}~\bibnamefont {Borsanyi}},
  \bibinfo {author} {\bibfnamefont {Z.}~\bibnamefont {Fodor}}, \bibinfo
  {author} {\bibfnamefont {J.}~\bibnamefont {G\"unther}}, \bibinfo {author}
  {\bibfnamefont {S.~D.}\ \bibnamefont {Katz}}, \bibinfo {author}
  {\bibfnamefont {C.}~\bibnamefont {Ratti}}, \ and\ \bibinfo {author}
  {\bibfnamefont {K.~K.}\ \bibnamefont {Szabo}},\ }\href {\doibase
  10.1016/j.physletb.2015.11.011} {\bibfield  {journal} {\bibinfo  {journal}
  {Phys. Lett. B}\ }\textbf {\bibinfo {volume} {751}},\ \bibinfo {pages} {559}
  (\bibinfo {year} {2015}{\natexlab{a}})},\ \Eprint
  {http://arxiv.org/abs/1507.07510} {arXiv:1507.07510 [hep-lat]} \BibitemShut
  {NoStop}%
\bibitem [{\citenamefont {Bazavov}\ \emph {et~al.}(2019)\citenamefont {Bazavov}
  \emph {et~al.}}]{Bazavov:2018mes}%
  \BibitemOpen
  \bibfield  {author} {\bibinfo {author} {\bibfnamefont {A.}~\bibnamefont
  {Bazavov}} \emph {et~al.} (\bibinfo {collaboration} {HotQCD}),\ }\href
  {\doibase 10.1016/j.physletb.2019.05.013} {\bibfield  {journal} {\bibinfo
  {journal} {Phys. Lett. B}\ }\textbf {\bibinfo {volume} {795}},\ \bibinfo
  {pages} {15} (\bibinfo {year} {2019})},\ \Eprint
  {http://arxiv.org/abs/1812.08235} {arXiv:1812.08235 [hep-lat]} \BibitemShut
  {NoStop}%
\bibitem [{\citenamefont {Borsanyi}\ \emph {et~al.}(2020)\citenamefont
  {Borsanyi}, \citenamefont {Fodor}, \citenamefont {Guenther}, \citenamefont
  {Kara}, \citenamefont {Katz}, \citenamefont {Parotto}, \citenamefont
  {Pasztor}, \citenamefont {Ratti},\ and\ \citenamefont
  {Szabo}}]{Borsanyi:2020fev}%
  \BibitemOpen
  \bibfield  {author} {\bibinfo {author} {\bibfnamefont {S.}~\bibnamefont
  {Borsanyi}}, \bibinfo {author} {\bibfnamefont {Z.}~\bibnamefont {Fodor}},
  \bibinfo {author} {\bibfnamefont {J.~N.}\ \bibnamefont {Guenther}}, \bibinfo
  {author} {\bibfnamefont {R.}~\bibnamefont {Kara}}, \bibinfo {author}
  {\bibfnamefont {S.~D.}\ \bibnamefont {Katz}}, \bibinfo {author}
  {\bibfnamefont {P.}~\bibnamefont {Parotto}}, \bibinfo {author} {\bibfnamefont
  {A.}~\bibnamefont {Pasztor}}, \bibinfo {author} {\bibfnamefont
  {C.}~\bibnamefont {Ratti}}, \ and\ \bibinfo {author} {\bibfnamefont {K.~K.}\
  \bibnamefont {Szabo}},\ }\href {\doibase 10.1103/PhysRevLett.125.052001}
  {\bibfield  {journal} {\bibinfo  {journal} {Phys. Rev. Lett.}\ }\textbf
  {\bibinfo {volume} {125}},\ \bibinfo {pages} {052001} (\bibinfo {year}
  {2020})},\ \Eprint {http://arxiv.org/abs/2002.02821} {arXiv:2002.02821
  [hep-lat]} \BibitemShut {NoStop}%
\bibitem [{\citenamefont {Borsanyi}\ \emph {et~al.}(2012)\citenamefont
  {Borsanyi}, \citenamefont {Fodor}, \citenamefont {Katz}, \citenamefont
  {Krieg}, \citenamefont {Ratti},\ and\ \citenamefont
  {Szabo}}]{Borsanyi:2011sw}%
  \BibitemOpen
  \bibfield  {author} {\bibinfo {author} {\bibfnamefont {S.}~\bibnamefont
  {Borsanyi}}, \bibinfo {author} {\bibfnamefont {Z.}~\bibnamefont {Fodor}},
  \bibinfo {author} {\bibfnamefont {S.~D.}\ \bibnamefont {Katz}}, \bibinfo
  {author} {\bibfnamefont {S.}~\bibnamefont {Krieg}}, \bibinfo {author}
  {\bibfnamefont {C.}~\bibnamefont {Ratti}}, \ and\ \bibinfo {author}
  {\bibfnamefont {K.}~\bibnamefont {Szabo}},\ }\href {\doibase
  10.1007/JHEP01(2012)138} {\bibfield  {journal} {\bibinfo  {journal} {JHEP}\
  }\textbf {\bibinfo {volume} {01}},\ \bibinfo {pages} {138} (\bibinfo {year}
  {2012})},\ \Eprint {http://arxiv.org/abs/1112.4416} {arXiv:1112.4416
  [hep-lat]} \BibitemShut {NoStop}%
\bibitem [{\citenamefont {Borsanyi}\ \emph {et~al.}(2013)\citenamefont
  {Borsanyi}, \citenamefont {Fodor}, \citenamefont {Katz}, \citenamefont
  {Krieg}, \citenamefont {Ratti},\ and\ \citenamefont
  {Szabo}}]{Borsanyi:2013hza}%
  \BibitemOpen
  \bibfield  {author} {\bibinfo {author} {\bibfnamefont {S.}~\bibnamefont
  {Borsanyi}}, \bibinfo {author} {\bibfnamefont {Z.}~\bibnamefont {Fodor}},
  \bibinfo {author} {\bibfnamefont {S.~D.}\ \bibnamefont {Katz}}, \bibinfo
  {author} {\bibfnamefont {S.}~\bibnamefont {Krieg}}, \bibinfo {author}
  {\bibfnamefont {C.}~\bibnamefont {Ratti}}, \ and\ \bibinfo {author}
  {\bibfnamefont {K.~K.}\ \bibnamefont {Szabo}},\ }\href {\doibase
  10.1103/PhysRevLett.111.062005} {\bibfield  {journal} {\bibinfo  {journal}
  {Phys. Rev. Lett.}\ }\textbf {\bibinfo {volume} {111}},\ \bibinfo {pages}
  {062005} (\bibinfo {year} {2013})},\ \Eprint {http://arxiv.org/abs/1305.5161}
  {arXiv:1305.5161 [hep-lat]} \BibitemShut {NoStop}%
\bibitem [{\citenamefont {Bellwied}\ \emph {et~al.}(2013)\citenamefont
  {Bellwied}, \citenamefont {Borsanyi}, \citenamefont {Fodor}, \citenamefont
  {Katz},\ and\ \citenamefont {Ratti}}]{Bellwied:2013cta}%
  \BibitemOpen
  \bibfield  {author} {\bibinfo {author} {\bibfnamefont {R.}~\bibnamefont
  {Bellwied}}, \bibinfo {author} {\bibfnamefont {S.}~\bibnamefont {Borsanyi}},
  \bibinfo {author} {\bibfnamefont {Z.}~\bibnamefont {Fodor}}, \bibinfo
  {author} {\bibfnamefont {S.~D.}\ \bibnamefont {Katz}}, \ and\ \bibinfo
  {author} {\bibfnamefont {C.}~\bibnamefont {Ratti}},\ }\href {\doibase
  10.1103/PhysRevLett.111.202302} {\bibfield  {journal} {\bibinfo  {journal}
  {Phys. Rev. Lett.}\ }\textbf {\bibinfo {volume} {111}},\ \bibinfo {pages}
  {202302} (\bibinfo {year} {2013})},\ \Eprint {http://arxiv.org/abs/1305.6297}
  {arXiv:1305.6297 [hep-lat]} \BibitemShut {NoStop}%
\bibitem [{\citenamefont {Borsanyi}\ \emph
  {et~al.}(2014{\natexlab{b}})\citenamefont {Borsanyi}, \citenamefont {Fodor},
  \citenamefont {Katz}, \citenamefont {Krieg}, \citenamefont {Ratti},\ and\
  \citenamefont {Szabo}}]{Borsanyi:2014ewa}%
  \BibitemOpen
  \bibfield  {author} {\bibinfo {author} {\bibfnamefont {S.}~\bibnamefont
  {Borsanyi}}, \bibinfo {author} {\bibfnamefont {Z.}~\bibnamefont {Fodor}},
  \bibinfo {author} {\bibfnamefont {S.~D.}\ \bibnamefont {Katz}}, \bibinfo
  {author} {\bibfnamefont {S.}~\bibnamefont {Krieg}}, \bibinfo {author}
  {\bibfnamefont {C.}~\bibnamefont {Ratti}}, \ and\ \bibinfo {author}
  {\bibfnamefont {K.~K.}\ \bibnamefont {Szabo}},\ }\href {\doibase
  10.1103/PhysRevLett.113.052301} {\bibfield  {journal} {\bibinfo  {journal}
  {Phys. Rev. Lett.}\ }\textbf {\bibinfo {volume} {113}},\ \bibinfo {pages}
  {052301} (\bibinfo {year} {2014}{\natexlab{b}})},\ \Eprint
  {http://arxiv.org/abs/1403.4576} {arXiv:1403.4576 [hep-lat]} \BibitemShut
  {NoStop}%
\bibitem [{\citenamefont {Bellwied}\ \emph
  {et~al.}(2015{\natexlab{b}})\citenamefont {Bellwied}, \citenamefont
  {Borsanyi}, \citenamefont {Fodor}, \citenamefont {Katz}, \citenamefont
  {Pasztor}, \citenamefont {Ratti},\ and\ \citenamefont
  {Szabo}}]{Bellwied:2015lba}%
  \BibitemOpen
  \bibfield  {author} {\bibinfo {author} {\bibfnamefont {R.}~\bibnamefont
  {Bellwied}}, \bibinfo {author} {\bibfnamefont {S.}~\bibnamefont {Borsanyi}},
  \bibinfo {author} {\bibfnamefont {Z.}~\bibnamefont {Fodor}}, \bibinfo
  {author} {\bibfnamefont {S.}~\bibnamefont {Katz}}, \bibinfo {author}
  {\bibfnamefont {A.}~\bibnamefont {Pasztor}}, \bibinfo {author} {\bibfnamefont
  {C.}~\bibnamefont {Ratti}}, \ and\ \bibinfo {author} {\bibfnamefont
  {K.}~\bibnamefont {Szabo}},\ }\href {\doibase 10.1103/PhysRevD.92.114505}
  {\bibfield  {journal} {\bibinfo  {journal} {Phys. Rev. D}\ }\textbf {\bibinfo
  {volume} {92}},\ \bibinfo {pages} {114505} (\bibinfo {year}
  {2015}{\natexlab{b}})},\ \Eprint {http://arxiv.org/abs/1507.04627}
  {arXiv:1507.04627 [hep-lat]} \BibitemShut {NoStop}%
\bibitem [{\citenamefont {Noronha-Hostler}\ \emph {et~al.}(2016)\citenamefont
  {Noronha-Hostler}, \citenamefont {Bellwied}, \citenamefont {Gunther},
  \citenamefont {Parotto}, \citenamefont {Pasztor}, \citenamefont
  {Portillo~Vazquez},\ and\ \citenamefont {Ratti}}]{Noronha-Hostler:2016rpd}%
  \BibitemOpen
  \bibfield  {author} {\bibinfo {author} {\bibfnamefont {J.}~\bibnamefont
  {Noronha-Hostler}}, \bibinfo {author} {\bibfnamefont {R.}~\bibnamefont
  {Bellwied}}, \bibinfo {author} {\bibfnamefont {J.}~\bibnamefont {Gunther}},
  \bibinfo {author} {\bibfnamefont {P.}~\bibnamefont {Parotto}}, \bibinfo
  {author} {\bibfnamefont {A.}~\bibnamefont {Pasztor}}, \bibinfo {author}
  {\bibfnamefont {I.}~\bibnamefont {Portillo~Vazquez}}, \ and\ \bibinfo
  {author} {\bibfnamefont {C.}~\bibnamefont {Ratti}},\ }\href@noop {} {\
  (\bibinfo {year} {2016})},\ \Eprint {http://arxiv.org/abs/1607.02527}
  {arXiv:1607.02527 [hep-ph]} \BibitemShut {NoStop}%
\bibitem [{\citenamefont {Bazavov}\ \emph
  {et~al.}(2017{\natexlab{b}})\citenamefont {Bazavov} \emph
  {et~al.}}]{Bazavov:2017tot}%
  \BibitemOpen
  \bibfield  {author} {\bibinfo {author} {\bibfnamefont {A.}~\bibnamefont
  {Bazavov}} \emph {et~al.} (\bibinfo {collaboration} {HotQCD}),\ }\href
  {\doibase 10.1103/PhysRevD.96.074510} {\bibfield  {journal} {\bibinfo
  {journal} {Phys. Rev.}\ }\textbf {\bibinfo {volume} {D96}},\ \bibinfo {pages}
  {074510} (\bibinfo {year} {2017}{\natexlab{b}})},\ \Eprint
  {http://arxiv.org/abs/1708.04897} {arXiv:1708.04897 [hep-lat]} \BibitemShut
  {NoStop}%
\bibitem [{\citenamefont {Borsanyi}\ \emph {et~al.}(2018)\citenamefont
  {Borsanyi}, \citenamefont {Fodor}, \citenamefont {Guenther}, \citenamefont
  {Katz}, \citenamefont {Szabo}, \citenamefont {Pasztor}, \citenamefont
  {Portillo},\ and\ \citenamefont {Ratti}}]{Borsanyi:2018grb}%
  \BibitemOpen
  \bibfield  {author} {\bibinfo {author} {\bibfnamefont {S.}~\bibnamefont
  {Borsanyi}}, \bibinfo {author} {\bibfnamefont {Z.}~\bibnamefont {Fodor}},
  \bibinfo {author} {\bibfnamefont {J.~N.}\ \bibnamefont {Guenther}}, \bibinfo
  {author} {\bibfnamefont {S.~K.}\ \bibnamefont {Katz}}, \bibinfo {author}
  {\bibfnamefont {K.~K.}\ \bibnamefont {Szabo}}, \bibinfo {author}
  {\bibfnamefont {A.}~\bibnamefont {Pasztor}}, \bibinfo {author} {\bibfnamefont
  {I.}~\bibnamefont {Portillo}}, \ and\ \bibinfo {author} {\bibfnamefont
  {C.}~\bibnamefont {Ratti}},\ }\href {\doibase 10.1007/JHEP10(2018)205}
  {\bibfield  {journal} {\bibinfo  {journal} {JHEP}\ }\textbf {\bibinfo
  {volume} {10}},\ \bibinfo {pages} {205} (\bibinfo {year} {2018})},\ \Eprint
  {http://arxiv.org/abs/1805.04445} {arXiv:1805.04445 [hep-lat]} \BibitemShut
  {NoStop}%
\bibitem [{\citenamefont {Bazavov}\ \emph {et~al.}(2020)\citenamefont {Bazavov}
  \emph {et~al.}}]{Bazavov:2020bjn}%
  \BibitemOpen
  \bibfield  {author} {\bibinfo {author} {\bibfnamefont {A.}~\bibnamefont
  {Bazavov}} \emph {et~al.},\ }\href {\doibase 10.1103/PhysRevD.101.074502}
  {\bibfield  {journal} {\bibinfo  {journal} {Phys. Rev. D}\ }\textbf {\bibinfo
  {volume} {101}},\ \bibinfo {pages} {074502} (\bibinfo {year} {2020})},\
  \Eprint {http://arxiv.org/abs/2001.08530} {arXiv:2001.08530 [hep-lat]}
  \BibitemShut {NoStop}%
\bibitem [{\citenamefont {Bazavov}\ \emph
  {et~al.}(2012{\natexlab{b}})\citenamefont {Bazavov} \emph
  {et~al.}}]{Bazavov:2012jq}%
  \BibitemOpen
  \bibfield  {author} {\bibinfo {author} {\bibfnamefont {A.}~\bibnamefont
  {Bazavov}} \emph {et~al.} (\bibinfo {collaboration} {HotQCD}),\ }\href
  {\doibase 10.1103/PhysRevD.86.034509} {\bibfield  {journal} {\bibinfo
  {journal} {Phys. Rev. D}\ }\textbf {\bibinfo {volume} {86}},\ \bibinfo
  {pages} {034509} (\bibinfo {year} {2012}{\natexlab{b}})},\ \Eprint
  {http://arxiv.org/abs/1203.0784} {arXiv:1203.0784 [hep-lat]} \BibitemShut
  {NoStop}%
\bibitem [{\citenamefont {Vovchenko}\ \emph
  {et~al.}(2017{\natexlab{a}})\citenamefont {Vovchenko}, \citenamefont
  {Gorenstein},\ and\ \citenamefont {Stoecker}}]{Vovchenko:2016rkn}%
  \BibitemOpen
  \bibfield  {author} {\bibinfo {author} {\bibfnamefont {V.}~\bibnamefont
  {Vovchenko}}, \bibinfo {author} {\bibfnamefont {M.~I.}\ \bibnamefont
  {Gorenstein}}, \ and\ \bibinfo {author} {\bibfnamefont {H.}~\bibnamefont
  {Stoecker}},\ }\href {\doibase 10.1103/PhysRevLett.118.182301} {\bibfield
  {journal} {\bibinfo  {journal} {Phys. Rev. Lett.}\ }\textbf {\bibinfo
  {volume} {118}},\ \bibinfo {pages} {182301} (\bibinfo {year}
  {2017}{\natexlab{a}})},\ \Eprint {http://arxiv.org/abs/1609.03975}
  {arXiv:1609.03975 [hep-ph]} \BibitemShut {NoStop}%
\bibitem [{\citenamefont {Alba}\ \emph {et~al.}(2017)\citenamefont {Alba} \emph
  {et~al.}}]{Alba:2017mqu}%
  \BibitemOpen
  \bibfield  {author} {\bibinfo {author} {\bibfnamefont {P.}~\bibnamefont
  {Alba}} \emph {et~al.},\ }\href {\doibase 10.1103/PhysRevD.96.034517}
  {\bibfield  {journal} {\bibinfo  {journal} {Phys. Rev.}\ }\textbf {\bibinfo
  {volume} {D96}},\ \bibinfo {pages} {034517} (\bibinfo {year} {2017})},\
  \Eprint {http://arxiv.org/abs/1702.01113} {arXiv:1702.01113 [hep-lat]}
  \BibitemShut {NoStop}%
\bibitem [{\citenamefont {Karsch}\ \emph
  {et~al.}(2003{\natexlab{a}})\citenamefont {Karsch}, \citenamefont {Redlich},\
  and\ \citenamefont {Tawfik}}]{Karsch:2003vd}%
  \BibitemOpen
  \bibfield  {author} {\bibinfo {author} {\bibfnamefont {F.}~\bibnamefont
  {Karsch}}, \bibinfo {author} {\bibfnamefont {K.}~\bibnamefont {Redlich}}, \
  and\ \bibinfo {author} {\bibfnamefont {A.}~\bibnamefont {Tawfik}},\ }\href
  {\doibase 10.1140/epjc/s2003-01228-y} {\bibfield  {journal} {\bibinfo
  {journal} {Eur. Phys. J. C}\ }\textbf {\bibinfo {volume} {29}},\ \bibinfo
  {pages} {549} (\bibinfo {year} {2003}{\natexlab{a}})},\ \Eprint
  {http://arxiv.org/abs/hep-ph/0303108} {arXiv:hep-ph/0303108} \BibitemShut
  {NoStop}%
\bibitem [{\citenamefont {Karsch}\ \emph
  {et~al.}(2003{\natexlab{b}})\citenamefont {Karsch}, \citenamefont {Redlich},\
  and\ \citenamefont {Tawfik}}]{Karsch:2003zq}%
  \BibitemOpen
  \bibfield  {author} {\bibinfo {author} {\bibfnamefont {F.}~\bibnamefont
  {Karsch}}, \bibinfo {author} {\bibfnamefont {K.}~\bibnamefont {Redlich}}, \
  and\ \bibinfo {author} {\bibfnamefont {A.}~\bibnamefont {Tawfik}},\ }\href
  {\doibase 10.1016/j.physletb.2003.08.001} {\bibfield  {journal} {\bibinfo
  {journal} {Phys. Lett. B}\ }\textbf {\bibinfo {volume} {571}},\ \bibinfo
  {pages} {67} (\bibinfo {year} {2003}{\natexlab{b}})},\ \Eprint
  {http://arxiv.org/abs/hep-ph/0306208} {arXiv:hep-ph/0306208} \BibitemShut
  {NoStop}%
\bibitem [{\citenamefont {Tawfik}(2005)}]{Tawfik:2004sw}%
  \BibitemOpen
  \bibfield  {author} {\bibinfo {author} {\bibfnamefont {A.}~\bibnamefont
  {Tawfik}},\ }\href {\doibase 10.1103/PhysRevD.71.054502} {\bibfield
  {journal} {\bibinfo  {journal} {Phys. Rev. D}\ }\textbf {\bibinfo {volume}
  {71}},\ \bibinfo {pages} {054502} (\bibinfo {year} {2005})},\ \Eprint
  {http://arxiv.org/abs/hep-ph/0412336} {arXiv:hep-ph/0412336} \BibitemShut
  {NoStop}%
\bibitem [{\citenamefont {Huovinen}\ and\ \citenamefont
  {Petreczky}(2010)}]{Huovinen:2009yb}%
  \BibitemOpen
  \bibfield  {author} {\bibinfo {author} {\bibfnamefont {P.}~\bibnamefont
  {Huovinen}}\ and\ \bibinfo {author} {\bibfnamefont {P.}~\bibnamefont
  {Petreczky}},\ }\href {\doibase 10.1016/j.nuclphysa.2010.02.015} {\bibfield
  {journal} {\bibinfo  {journal} {Nucl. Phys. A}\ }\textbf {\bibinfo {volume}
  {837}},\ \bibinfo {pages} {26} (\bibinfo {year} {2010})},\ \Eprint
  {http://arxiv.org/abs/0912.2541} {arXiv:0912.2541 [hep-ph]} \BibitemShut
  {NoStop}%
\bibitem [{\citenamefont {Ratti}\ \emph {et~al.}(2011)\citenamefont {Ratti},
  \citenamefont {Borsanyi}, \citenamefont {Fodor}, \citenamefont {Hoelbling},
  \citenamefont {Katz}, \citenamefont {Krieg},\ and\ \citenamefont
  {Szabo}}]{Ratti:2010kj}%
  \BibitemOpen
  \bibfield  {author} {\bibinfo {author} {\bibfnamefont {C.}~\bibnamefont
  {Ratti}}, \bibinfo {author} {\bibfnamefont {S.}~\bibnamefont {Borsanyi}},
  \bibinfo {author} {\bibfnamefont {Z.}~\bibnamefont {Fodor}}, \bibinfo
  {author} {\bibfnamefont {C.}~\bibnamefont {Hoelbling}}, \bibinfo {author}
  {\bibfnamefont {S.~D.}\ \bibnamefont {Katz}}, \bibinfo {author}
  {\bibfnamefont {S.}~\bibnamefont {Krieg}}, \ and\ \bibinfo {author}
  {\bibfnamefont {K.~K.}\ \bibnamefont {Szabo}} (\bibinfo {collaboration}
  {Wuppertal-Budapest}),\ }\href {\doibase 10.1016/j.nuclphysa.2011.02.052}
  {\bibfield  {journal} {\bibinfo  {journal} {Nucl. Phys. A}\ }\textbf
  {\bibinfo {volume} {855}},\ \bibinfo {pages} {253} (\bibinfo {year}
  {2011})},\ \Eprint {http://arxiv.org/abs/1012.5215} {arXiv:1012.5215
  [hep-lat]} \BibitemShut {NoStop}%
\bibitem [{\citenamefont {Alba}\ \emph {et~al.}(2015)\citenamefont {Alba},
  \citenamefont {Bellwied}, \citenamefont {Bluhm}, \citenamefont
  {Mantovani~Sarti}, \citenamefont {Nahrgang},\ and\ \citenamefont
  {Ratti}}]{Alba:2015iva}%
  \BibitemOpen
  \bibfield  {author} {\bibinfo {author} {\bibfnamefont {P.}~\bibnamefont
  {Alba}}, \bibinfo {author} {\bibfnamefont {R.}~\bibnamefont {Bellwied}},
  \bibinfo {author} {\bibfnamefont {M.}~\bibnamefont {Bluhm}}, \bibinfo
  {author} {\bibfnamefont {V.}~\bibnamefont {Mantovani~Sarti}}, \bibinfo
  {author} {\bibfnamefont {M.}~\bibnamefont {Nahrgang}}, \ and\ \bibinfo
  {author} {\bibfnamefont {C.}~\bibnamefont {Ratti}},\ }\href {\doibase
  10.1103/PhysRevC.92.064910} {\bibfield  {journal} {\bibinfo  {journal} {Phys.
  Rev. C}\ }\textbf {\bibinfo {volume} {92}},\ \bibinfo {pages} {064910}
  (\bibinfo {year} {2015})},\ \Eprint {http://arxiv.org/abs/1504.03262}
  {arXiv:1504.03262 [hep-ph]} \BibitemShut {NoStop}%
\bibitem [{\citenamefont {Huovinen}\ and\ \citenamefont
  {Petreczky}(2018)}]{Huovinen:2017ogf}%
  \BibitemOpen
  \bibfield  {author} {\bibinfo {author} {\bibfnamefont {P.}~\bibnamefont
  {Huovinen}}\ and\ \bibinfo {author} {\bibfnamefont {P.}~\bibnamefont
  {Petreczky}},\ }\href {\doibase 10.1016/j.physletb.2017.12.001} {\bibfield
  {journal} {\bibinfo  {journal} {Phys. Lett. B}\ }\textbf {\bibinfo {volume}
  {777}},\ \bibinfo {pages} {125} (\bibinfo {year} {2018})},\ \Eprint
  {http://arxiv.org/abs/1708.00879} {arXiv:1708.00879 [hep-ph]} \BibitemShut
  {NoStop}%
\bibitem [{\citenamefont {Bellwied}\ \emph
  {et~al.}(2019{\natexlab{a}})\citenamefont {Bellwied}, \citenamefont
  {Borsanyi}, \citenamefont {Fodor}, \citenamefont {Guenther}, \citenamefont
  {Noronha-Hostler}, \citenamefont {Parotto}, \citenamefont {Pasztor},
  \citenamefont {Ratti},\ and\ \citenamefont {Stafford}}]{Bellwied:2019pxh}%
  \BibitemOpen
  \bibfield  {author} {\bibinfo {author} {\bibfnamefont {R.}~\bibnamefont
  {Bellwied}}, \bibinfo {author} {\bibfnamefont {S.}~\bibnamefont {Borsanyi}},
  \bibinfo {author} {\bibfnamefont {Z.}~\bibnamefont {Fodor}}, \bibinfo
  {author} {\bibfnamefont {J.~N.}\ \bibnamefont {Guenther}}, \bibinfo {author}
  {\bibfnamefont {J.}~\bibnamefont {Noronha-Hostler}}, \bibinfo {author}
  {\bibfnamefont {P.}~\bibnamefont {Parotto}}, \bibinfo {author} {\bibfnamefont
  {A.}~\bibnamefont {Pasztor}}, \bibinfo {author} {\bibfnamefont
  {C.}~\bibnamefont {Ratti}}, \ and\ \bibinfo {author} {\bibfnamefont {J.~M.}\
  \bibnamefont {Stafford}},\ }\href@noop {} {\  (\bibinfo {year}
  {2019}{\natexlab{a}})},\ \Eprint {http://arxiv.org/abs/1910.14592}
  {arXiv:1910.14592 [hep-lat]} \BibitemShut {NoStop}%
\bibitem [{\citenamefont {Vovchenko}\ \emph {et~al.}(2015)\citenamefont
  {Vovchenko}, \citenamefont {Anchishkin},\ and\ \citenamefont
  {Gorenstein}}]{Vovchenko:2014pka}%
  \BibitemOpen
  \bibfield  {author} {\bibinfo {author} {\bibfnamefont {V.}~\bibnamefont
  {Vovchenko}}, \bibinfo {author} {\bibfnamefont {D.~V.}\ \bibnamefont
  {Anchishkin}}, \ and\ \bibinfo {author} {\bibfnamefont {M.~I.}\ \bibnamefont
  {Gorenstein}},\ }\href {\doibase 10.1103/PhysRevC.91.024905} {\bibfield
  {journal} {\bibinfo  {journal} {Phys. Rev. C}\ }\textbf {\bibinfo {volume}
  {91}},\ \bibinfo {pages} {024905} (\bibinfo {year} {2015})},\ \Eprint
  {http://arxiv.org/abs/1412.5478} {arXiv:1412.5478 [nucl-th]} \BibitemShut
  {NoStop}%
\bibitem [{\citenamefont {Becattini}\ \emph {et~al.}(2013)\citenamefont
  {Becattini}, \citenamefont {Bleicher}, \citenamefont {Kollegger},
  \citenamefont {Schuster}, \citenamefont {Steinheimer},\ and\ \citenamefont
  {Stock}}]{Becattini:2012xb}%
  \BibitemOpen
  \bibfield  {author} {\bibinfo {author} {\bibfnamefont {F.}~\bibnamefont
  {Becattini}}, \bibinfo {author} {\bibfnamefont {M.}~\bibnamefont {Bleicher}},
  \bibinfo {author} {\bibfnamefont {T.}~\bibnamefont {Kollegger}}, \bibinfo
  {author} {\bibfnamefont {T.}~\bibnamefont {Schuster}}, \bibinfo {author}
  {\bibfnamefont {J.}~\bibnamefont {Steinheimer}}, \ and\ \bibinfo {author}
  {\bibfnamefont {R.}~\bibnamefont {Stock}},\ }\href {\doibase
  10.1103/PhysRevLett.111.082302} {\bibfield  {journal} {\bibinfo  {journal}
  {Phys. Rev. Lett.}\ }\textbf {\bibinfo {volume} {111}},\ \bibinfo {pages}
  {082302} (\bibinfo {year} {2013})},\ \Eprint {http://arxiv.org/abs/1212.2431}
  {arXiv:1212.2431 [nucl-th]} \BibitemShut {NoStop}%
\bibitem [{\citenamefont {Alba}\ \emph {et~al.}(2014)\citenamefont {Alba},
  \citenamefont {Alberico}, \citenamefont {Bellwied}, \citenamefont {Bluhm},
  \citenamefont {Mantovani~Sarti}, \citenamefont {Nahrgang},\ and\
  \citenamefont {Ratti}}]{Alba:2014eba}%
  \BibitemOpen
  \bibfield  {author} {\bibinfo {author} {\bibfnamefont {P.}~\bibnamefont
  {Alba}}, \bibinfo {author} {\bibfnamefont {W.}~\bibnamefont {Alberico}},
  \bibinfo {author} {\bibfnamefont {R.}~\bibnamefont {Bellwied}}, \bibinfo
  {author} {\bibfnamefont {M.}~\bibnamefont {Bluhm}}, \bibinfo {author}
  {\bibfnamefont {V.}~\bibnamefont {Mantovani~Sarti}}, \bibinfo {author}
  {\bibfnamefont {M.}~\bibnamefont {Nahrgang}}, \ and\ \bibinfo {author}
  {\bibfnamefont {C.}~\bibnamefont {Ratti}},\ }\href {\doibase
  10.1016/j.physletb.2014.09.052} {\bibfield  {journal} {\bibinfo  {journal}
  {Phys. Lett. B}\ }\textbf {\bibinfo {volume} {738}},\ \bibinfo {pages} {305}
  (\bibinfo {year} {2014})},\ \Eprint {http://arxiv.org/abs/1403.4903}
  {arXiv:1403.4903 [hep-ph]} \BibitemShut {NoStop}%
\bibitem [{\citenamefont {Vovchenko}\ \emph {et~al.}(2016)\citenamefont
  {Vovchenko}, \citenamefont {Begun},\ and\ \citenamefont
  {Gorenstein}}]{Vovchenko:2015idt}%
  \BibitemOpen
  \bibfield  {author} {\bibinfo {author} {\bibfnamefont {V.}~\bibnamefont
  {Vovchenko}}, \bibinfo {author} {\bibfnamefont {V.~V.}\ \bibnamefont
  {Begun}}, \ and\ \bibinfo {author} {\bibfnamefont {M.~I.}\ \bibnamefont
  {Gorenstein}},\ }\href {\doibase 10.1103/PhysRevC.93.064906} {\bibfield
  {journal} {\bibinfo  {journal} {Phys. Rev. C}\ }\textbf {\bibinfo {volume}
  {93}},\ \bibinfo {pages} {064906} (\bibinfo {year} {2016})},\ \Eprint
  {http://arxiv.org/abs/1512.08025} {arXiv:1512.08025 [nucl-th]} \BibitemShut
  {NoStop}%
\bibitem [{\citenamefont {Andronic}\ \emph {et~al.}(2018)\citenamefont
  {Andronic}, \citenamefont {Braun-Munzinger}, \citenamefont {Redlich},\ and\
  \citenamefont {Stachel}}]{Andronic:2017pug}%
  \BibitemOpen
  \bibfield  {author} {\bibinfo {author} {\bibfnamefont {A.}~\bibnamefont
  {Andronic}}, \bibinfo {author} {\bibfnamefont {P.}~\bibnamefont
  {Braun-Munzinger}}, \bibinfo {author} {\bibfnamefont {K.}~\bibnamefont
  {Redlich}}, \ and\ \bibinfo {author} {\bibfnamefont {J.}~\bibnamefont
  {Stachel}},\ }\href {\doibase 10.1038/s41586-018-0491-6} {\bibfield
  {journal} {\bibinfo  {journal} {Nature}\ }\textbf {\bibinfo {volume} {561}},\
  \bibinfo {pages} {321} (\bibinfo {year} {2018})},\ \Eprint
  {http://arxiv.org/abs/1710.09425} {arXiv:1710.09425 [nucl-th]} \BibitemShut
  {NoStop}%
\bibitem [{\citenamefont {Bellwied}\ \emph
  {et~al.}(2019{\natexlab{b}})\citenamefont {Bellwied}, \citenamefont
  {Noronha-Hostler}, \citenamefont {Parotto}, \citenamefont {Portillo~Vazquez},
  \citenamefont {Ratti},\ and\ \citenamefont {Stafford}}]{Bellwied:2018tkc}%
  \BibitemOpen
  \bibfield  {author} {\bibinfo {author} {\bibfnamefont {R.}~\bibnamefont
  {Bellwied}}, \bibinfo {author} {\bibfnamefont {J.}~\bibnamefont
  {Noronha-Hostler}}, \bibinfo {author} {\bibfnamefont {P.}~\bibnamefont
  {Parotto}}, \bibinfo {author} {\bibfnamefont {I.}~\bibnamefont
  {Portillo~Vazquez}}, \bibinfo {author} {\bibfnamefont {C.}~\bibnamefont
  {Ratti}}, \ and\ \bibinfo {author} {\bibfnamefont {J.~M.}\ \bibnamefont
  {Stafford}},\ }\href {\doibase 10.1103/PhysRevC.99.034912} {\bibfield
  {journal} {\bibinfo  {journal} {Phys. Rev.}\ }\textbf {\bibinfo {volume}
  {C99}},\ \bibinfo {pages} {034912} (\bibinfo {year} {2019}{\natexlab{b}})},\
  \Eprint {http://arxiv.org/abs/1805.00088} {arXiv:1805.00088 [hep-ph]}
  \BibitemShut {NoStop}%
\bibitem [{\citenamefont {Alba}\ \emph {et~al.}(2020)\citenamefont {Alba},
  \citenamefont {Sarti}, \citenamefont {Noronha-Hostler}, \citenamefont
  {Parotto}, \citenamefont {Portillo-Vazquez}, \citenamefont {Ratti},\ and\
  \citenamefont {Stafford}}]{Alba:2020jir}%
  \BibitemOpen
  \bibfield  {author} {\bibinfo {author} {\bibfnamefont {P.}~\bibnamefont
  {Alba}}, \bibinfo {author} {\bibfnamefont {V.~M.}\ \bibnamefont {Sarti}},
  \bibinfo {author} {\bibfnamefont {J.}~\bibnamefont {Noronha-Hostler}},
  \bibinfo {author} {\bibfnamefont {P.}~\bibnamefont {Parotto}}, \bibinfo
  {author} {\bibfnamefont {I.}~\bibnamefont {Portillo-Vazquez}}, \bibinfo
  {author} {\bibfnamefont {C.}~\bibnamefont {Ratti}}, \ and\ \bibinfo {author}
  {\bibfnamefont {J.~M.}\ \bibnamefont {Stafford}},\ }\href {\doibase
  10.1103/PhysRevC.101.054905} {\bibfield  {journal} {\bibinfo  {journal}
  {Phys. Rev. C}\ }\textbf {\bibinfo {volume} {101}},\ \bibinfo {pages}
  {054905} (\bibinfo {year} {2020})},\ \Eprint
  {http://arxiv.org/abs/2002.12395} {arXiv:2002.12395 [hep-ph]} \BibitemShut
  {NoStop}%
\bibitem [{\citenamefont {Bluhm}\ and\ \citenamefont
  {Nahrgang}(2019)}]{Bluhm:2018aei}%
  \BibitemOpen
  \bibfield  {author} {\bibinfo {author} {\bibfnamefont {M.}~\bibnamefont
  {Bluhm}}\ and\ \bibinfo {author} {\bibfnamefont {M.}~\bibnamefont
  {Nahrgang}},\ }\href {\doibase 10.1140/epjc/s10052-019-6661-3} {\bibfield
  {journal} {\bibinfo  {journal} {Eur. Phys. J.}\ }\textbf {\bibinfo {volume}
  {C79}},\ \bibinfo {pages} {155} (\bibinfo {year} {2019})},\ \Eprint
  {http://arxiv.org/abs/1806.04499} {arXiv:1806.04499 [nucl-th]} \BibitemShut
  {NoStop}%
\bibitem [{\citenamefont {Flor}\ \emph {et~al.}(2021)\citenamefont {Flor},
  \citenamefont {Olinger},\ and\ \citenamefont {Bellwied}}]{Flor:2020fdw}%
  \BibitemOpen
  \bibfield  {author} {\bibinfo {author} {\bibfnamefont {F.~A.}\ \bibnamefont
  {Flor}}, \bibinfo {author} {\bibfnamefont {G.}~\bibnamefont {Olinger}}, \
  and\ \bibinfo {author} {\bibfnamefont {R.}~\bibnamefont {Bellwied}},\ }\href
  {\doibase 10.1016/j.physletb.2021.136098} {\bibfield  {journal} {\bibinfo
  {journal} {Phys. Lett. B}\ }\textbf {\bibinfo {volume} {814}},\ \bibinfo
  {pages} {136098} (\bibinfo {year} {2021})},\ \Eprint
  {http://arxiv.org/abs/2009.14781} {arXiv:2009.14781 [nucl-ex]} \BibitemShut
  {NoStop}%
\bibitem [{\citenamefont {Bazavov}\ \emph
  {et~al.}(2014{\natexlab{b}})\citenamefont {Bazavov} \emph
  {et~al.}}]{Bazavov:2014xya}%
  \BibitemOpen
  \bibfield  {author} {\bibinfo {author} {\bibfnamefont {A.}~\bibnamefont
  {Bazavov}} \emph {et~al.},\ }\href {\doibase 10.1103/PhysRevLett.113.072001}
  {\bibfield  {journal} {\bibinfo  {journal} {Phys. Rev. Lett.}\ }\textbf
  {\bibinfo {volume} {113}},\ \bibinfo {pages} {072001} (\bibinfo {year}
  {2014}{\natexlab{b}})},\ \Eprint {http://arxiv.org/abs/1404.6511}
  {arXiv:1404.6511 [hep-lat]} \BibitemShut {NoStop}%
\bibitem [{\citenamefont {Bellwied}\ \emph {et~al.}(2021)\citenamefont
  {Bellwied}, \citenamefont {Borsanyi}, \citenamefont {Fodor}, \citenamefont
  {Guenther}, \citenamefont {Katz}, \citenamefont {Parotto}, \citenamefont
  {Pasztor}, \citenamefont {Pesznyak}, \citenamefont {Ratti},\ and\
  \citenamefont {Szabo}}]{Bellwied:2021nrt}%
  \BibitemOpen
  \bibfield  {author} {\bibinfo {author} {\bibfnamefont {R.}~\bibnamefont
  {Bellwied}}, \bibinfo {author} {\bibfnamefont {S.}~\bibnamefont {Borsanyi}},
  \bibinfo {author} {\bibfnamefont {Z.}~\bibnamefont {Fodor}}, \bibinfo
  {author} {\bibfnamefont {J.~N.}\ \bibnamefont {Guenther}}, \bibinfo {author}
  {\bibfnamefont {S.~D.}\ \bibnamefont {Katz}}, \bibinfo {author}
  {\bibfnamefont {P.}~\bibnamefont {Parotto}}, \bibinfo {author} {\bibfnamefont
  {A.}~\bibnamefont {Pasztor}}, \bibinfo {author} {\bibfnamefont
  {D.}~\bibnamefont {Pesznyak}}, \bibinfo {author} {\bibfnamefont
  {C.}~\bibnamefont {Ratti}}, \ and\ \bibinfo {author} {\bibfnamefont {K.~K.}\
  \bibnamefont {Szabo}},\ }\href@noop {} {\  (\bibinfo {year} {2021})},\
  \Eprint {http://arxiv.org/abs/2102.06625} {arXiv:2102.06625 [hep-lat]}
  \BibitemShut {NoStop}%
\bibitem [{\citenamefont {Yen}\ \emph {et~al.}(1997)\citenamefont {Yen},
  \citenamefont {Gorenstein}, \citenamefont {Greiner},\ and\ \citenamefont
  {Yang}}]{Yen:1997rv}%
  \BibitemOpen
  \bibfield  {author} {\bibinfo {author} {\bibfnamefont {G.~D.}\ \bibnamefont
  {Yen}}, \bibinfo {author} {\bibfnamefont {M.~I.}\ \bibnamefont {Gorenstein}},
  \bibinfo {author} {\bibfnamefont {W.}~\bibnamefont {Greiner}}, \ and\
  \bibinfo {author} {\bibfnamefont {S.-N.}\ \bibnamefont {Yang}},\ }\href
  {\doibase 10.1103/PhysRevC.56.2210} {\bibfield  {journal} {\bibinfo
  {journal} {Phys. Rev. C}\ }\textbf {\bibinfo {volume} {56}},\ \bibinfo
  {pages} {2210} (\bibinfo {year} {1997})},\ \Eprint
  {http://arxiv.org/abs/nucl-th/9711062} {arXiv:nucl-th/9711062} \BibitemShut
  {NoStop}%
\bibitem [{\citenamefont {Andronic}\ \emph {et~al.}(2012)\citenamefont
  {Andronic}, \citenamefont {Braun-Munzinger}, \citenamefont {Stachel},\ and\
  \citenamefont {Winn}}]{Andronic:2012ut}%
  \BibitemOpen
  \bibfield  {author} {\bibinfo {author} {\bibfnamefont {A.}~\bibnamefont
  {Andronic}}, \bibinfo {author} {\bibfnamefont {P.}~\bibnamefont
  {Braun-Munzinger}}, \bibinfo {author} {\bibfnamefont {J.}~\bibnamefont
  {Stachel}}, \ and\ \bibinfo {author} {\bibfnamefont {M.}~\bibnamefont
  {Winn}},\ }\href {\doibase 10.1016/j.physletb.2012.10.001} {\bibfield
  {journal} {\bibinfo  {journal} {Phys. Lett. B}\ }\textbf {\bibinfo {volume}
  {718}},\ \bibinfo {pages} {80} (\bibinfo {year} {2012})},\ \Eprint
  {http://arxiv.org/abs/1201.0693} {arXiv:1201.0693 [nucl-th]} \BibitemShut
  {NoStop}%
\bibitem [{\citenamefont {Noronha-Hostler}\ \emph {et~al.}(2012)\citenamefont
  {Noronha-Hostler}, \citenamefont {Noronha},\ and\ \citenamefont
  {Greiner}}]{NoronhaHostler:2012ug}%
  \BibitemOpen
  \bibfield  {author} {\bibinfo {author} {\bibfnamefont {J.}~\bibnamefont
  {Noronha-Hostler}}, \bibinfo {author} {\bibfnamefont {J.}~\bibnamefont
  {Noronha}}, \ and\ \bibinfo {author} {\bibfnamefont {C.}~\bibnamefont
  {Greiner}},\ }\href {\doibase 10.1103/PhysRevC.86.024913} {\bibfield
  {journal} {\bibinfo  {journal} {Phys. Rev.}\ }\textbf {\bibinfo {volume}
  {C86}},\ \bibinfo {pages} {024913} (\bibinfo {year} {2012})},\ \Eprint
  {http://arxiv.org/abs/1206.5138} {arXiv:1206.5138 [nucl-th]} \BibitemShut
  {NoStop}%
\bibitem [{\citenamefont {Bhattacharyya}\ \emph {et~al.}(2014)\citenamefont
  {Bhattacharyya}, \citenamefont {Das}, \citenamefont {Ghosh}, \citenamefont
  {Ray},\ and\ \citenamefont {Samanta}}]{Bhattacharyya:2013oya}%
  \BibitemOpen
  \bibfield  {author} {\bibinfo {author} {\bibfnamefont {A.}~\bibnamefont
  {Bhattacharyya}}, \bibinfo {author} {\bibfnamefont {S.}~\bibnamefont {Das}},
  \bibinfo {author} {\bibfnamefont {S.~K.}\ \bibnamefont {Ghosh}}, \bibinfo
  {author} {\bibfnamefont {R.}~\bibnamefont {Ray}}, \ and\ \bibinfo {author}
  {\bibfnamefont {S.}~\bibnamefont {Samanta}},\ }\href {\doibase
  10.1103/PhysRevC.90.034909} {\bibfield  {journal} {\bibinfo  {journal} {Phys.
  Rev. C}\ }\textbf {\bibinfo {volume} {90}},\ \bibinfo {pages} {034909}
  (\bibinfo {year} {2014})},\ \Eprint {http://arxiv.org/abs/1310.2793}
  {arXiv:1310.2793 [hep-ph]} \BibitemShut {NoStop}%
\bibitem [{\citenamefont {Albright}\ \emph {et~al.}(2015)\citenamefont
  {Albright}, \citenamefont {Kapusta},\ and\ \citenamefont
  {Young}}]{Albright:2015uua}%
  \BibitemOpen
  \bibfield  {author} {\bibinfo {author} {\bibfnamefont {M.}~\bibnamefont
  {Albright}}, \bibinfo {author} {\bibfnamefont {J.}~\bibnamefont {Kapusta}}, \
  and\ \bibinfo {author} {\bibfnamefont {C.}~\bibnamefont {Young}},\ }\href
  {\doibase 10.1103/PhysRevC.92.044904} {\bibfield  {journal} {\bibinfo
  {journal} {Phys. Rev. C}\ }\textbf {\bibinfo {volume} {92}},\ \bibinfo
  {pages} {044904} (\bibinfo {year} {2015})},\ \Eprint
  {http://arxiv.org/abs/1506.03408} {arXiv:1506.03408 [nucl-th]} \BibitemShut
  {NoStop}%
\bibitem [{\citenamefont {Satarov}\ \emph {et~al.}(2017)\citenamefont
  {Satarov}, \citenamefont {Vovchenko}, \citenamefont {Alba}, \citenamefont
  {Gorenstein},\ and\ \citenamefont {Stoecker}}]{Satarov:2016peb}%
  \BibitemOpen
  \bibfield  {author} {\bibinfo {author} {\bibfnamefont {L.~M.}\ \bibnamefont
  {Satarov}}, \bibinfo {author} {\bibfnamefont {V.}~\bibnamefont {Vovchenko}},
  \bibinfo {author} {\bibfnamefont {P.}~\bibnamefont {Alba}}, \bibinfo {author}
  {\bibfnamefont {M.~I.}\ \bibnamefont {Gorenstein}}, \ and\ \bibinfo {author}
  {\bibfnamefont {H.}~\bibnamefont {Stoecker}},\ }\href {\doibase
  10.1103/PhysRevC.95.024902} {\bibfield  {journal} {\bibinfo  {journal} {Phys.
  Rev. C}\ }\textbf {\bibinfo {volume} {95}},\ \bibinfo {pages} {024902}
  (\bibinfo {year} {2017})},\ \Eprint {http://arxiv.org/abs/1610.08753}
  {arXiv:1610.08753 [nucl-th]} \BibitemShut {NoStop}%
\bibitem [{\citenamefont {Vovchenko}\ \emph
  {et~al.}(2017{\natexlab{b}})\citenamefont {Vovchenko}, \citenamefont
  {Pasztor}, \citenamefont {Fodor}, \citenamefont {Katz},\ and\ \citenamefont
  {Stoecker}}]{Vovchenko:2017xad}%
  \BibitemOpen
  \bibfield  {author} {\bibinfo {author} {\bibfnamefont {V.}~\bibnamefont
  {Vovchenko}}, \bibinfo {author} {\bibfnamefont {A.}~\bibnamefont {Pasztor}},
  \bibinfo {author} {\bibfnamefont {Z.}~\bibnamefont {Fodor}}, \bibinfo
  {author} {\bibfnamefont {S.~D.}\ \bibnamefont {Katz}}, \ and\ \bibinfo
  {author} {\bibfnamefont {H.}~\bibnamefont {Stoecker}},\ }\href {\doibase
  10.1016/j.physletb.2017.10.042} {\bibfield  {journal} {\bibinfo  {journal}
  {Phys. Lett. B}\ }\textbf {\bibinfo {volume} {775}},\ \bibinfo {pages} {71}
  (\bibinfo {year} {2017}{\natexlab{b}})},\ \Eprint
  {http://arxiv.org/abs/1708.02852} {arXiv:1708.02852 [hep-ph]} \BibitemShut
  {NoStop}%
\bibitem [{\citenamefont {Alba}\ and\ \citenamefont
  {Oliva}(2019)}]{Alba:2017bbr}%
  \BibitemOpen
  \bibfield  {author} {\bibinfo {author} {\bibfnamefont {P.}~\bibnamefont
  {Alba}}\ and\ \bibinfo {author} {\bibfnamefont {L.}~\bibnamefont {Oliva}},\
  }\href {\doibase 10.1103/PhysRevC.99.055207} {\bibfield  {journal} {\bibinfo
  {journal} {Phys. Rev. C}\ }\textbf {\bibinfo {volume} {99}},\ \bibinfo
  {pages} {055207} (\bibinfo {year} {2019})},\ \Eprint
  {http://arxiv.org/abs/1711.02797} {arXiv:1711.02797 [nucl-th]} \BibitemShut
  {NoStop}%
\bibitem [{\citenamefont {Vovchenko}\ \emph {et~al.}(2019)\citenamefont
  {Vovchenko}, \citenamefont {Gorenstein}, \citenamefont {Greiner},\ and\
  \citenamefont {Stoecker}}]{Vovchenko:2018eod}%
  \BibitemOpen
  \bibfield  {author} {\bibinfo {author} {\bibfnamefont {V.}~\bibnamefont
  {Vovchenko}}, \bibinfo {author} {\bibfnamefont {M.~I.}\ \bibnamefont
  {Gorenstein}}, \bibinfo {author} {\bibfnamefont {C.}~\bibnamefont {Greiner}},
  \ and\ \bibinfo {author} {\bibfnamefont {H.}~\bibnamefont {Stoecker}},\
  }\href {\doibase 10.1103/PhysRevC.99.045204} {\bibfield  {journal} {\bibinfo
  {journal} {Phys. Rev. C}\ }\textbf {\bibinfo {volume} {99}},\ \bibinfo
  {pages} {045204} (\bibinfo {year} {2019})},\ \Eprint
  {http://arxiv.org/abs/1811.05737} {arXiv:1811.05737 [hep-ph]} \BibitemShut
  {NoStop}%
\bibitem [{\citenamefont {Motornenko}\ \emph {et~al.}(2020)\citenamefont
  {Motornenko}, \citenamefont {Pal}, \citenamefont {Bhattacharyya},
  \citenamefont {Steinheimer},\ and\ \citenamefont
  {Stoecker}}]{Motornenko:2020yme}%
  \BibitemOpen
  \bibfield  {author} {\bibinfo {author} {\bibfnamefont {A.}~\bibnamefont
  {Motornenko}}, \bibinfo {author} {\bibfnamefont {S.}~\bibnamefont {Pal}},
  \bibinfo {author} {\bibfnamefont {A.}~\bibnamefont {Bhattacharyya}}, \bibinfo
  {author} {\bibfnamefont {J.}~\bibnamefont {Steinheimer}}, \ and\ \bibinfo
  {author} {\bibfnamefont {H.}~\bibnamefont {Stoecker}},\ }\href@noop {} {\
  (\bibinfo {year} {2020})},\ \Eprint {http://arxiv.org/abs/2009.10848}
  {arXiv:2009.10848 [hep-ph]} \BibitemShut {NoStop}%
\bibitem [{\citenamefont {Vovchenko}\ \emph
  {et~al.}(2017{\natexlab{c}})\citenamefont {Vovchenko}, \citenamefont
  {Motornenko}, \citenamefont {Alba}, \citenamefont {Gorenstein}, \citenamefont
  {Satarov},\ and\ \citenamefont {Stoecker}}]{Vovchenko:2017zpj}%
  \BibitemOpen
  \bibfield  {author} {\bibinfo {author} {\bibfnamefont {V.}~\bibnamefont
  {Vovchenko}}, \bibinfo {author} {\bibfnamefont {A.}~\bibnamefont
  {Motornenko}}, \bibinfo {author} {\bibfnamefont {P.}~\bibnamefont {Alba}},
  \bibinfo {author} {\bibfnamefont {M.~I.}\ \bibnamefont {Gorenstein}},
  \bibinfo {author} {\bibfnamefont {L.~M.}\ \bibnamefont {Satarov}}, \ and\
  \bibinfo {author} {\bibfnamefont {H.}~\bibnamefont {Stoecker}},\ }\href
  {\doibase 10.1103/PhysRevC.96.045202} {\bibfield  {journal} {\bibinfo
  {journal} {Phys. Rev. C}\ }\textbf {\bibinfo {volume} {96}},\ \bibinfo
  {pages} {045202} (\bibinfo {year} {2017}{\natexlab{c}})},\ \Eprint
  {http://arxiv.org/abs/1707.09215} {arXiv:1707.09215 [nucl-th]} \BibitemShut
  {NoStop}%
\bibitem [{\citenamefont {Samanta}\ and\ \citenamefont
  {Mohanty}(2018)}]{Samanta:2017yhh}%
  \BibitemOpen
  \bibfield  {author} {\bibinfo {author} {\bibfnamefont {S.}~\bibnamefont
  {Samanta}}\ and\ \bibinfo {author} {\bibfnamefont {B.}~\bibnamefont
  {Mohanty}},\ }\href {\doibase 10.1103/PhysRevC.97.015201} {\bibfield
  {journal} {\bibinfo  {journal} {Phys. Rev. C}\ }\textbf {\bibinfo {volume}
  {97}},\ \bibinfo {pages} {015201} (\bibinfo {year} {2018})},\ \Eprint
  {http://arxiv.org/abs/1709.04446} {arXiv:1709.04446 [hep-ph]} \BibitemShut
  {NoStop}%
\bibitem [{\citenamefont {Sarkar}\ and\ \citenamefont
  {Ghosh}(2018)}]{Sarkar:2018mbk}%
  \BibitemOpen
  \bibfield  {author} {\bibinfo {author} {\bibfnamefont {N.}~\bibnamefont
  {Sarkar}}\ and\ \bibinfo {author} {\bibfnamefont {P.}~\bibnamefont {Ghosh}},\
  }\href {\doibase 10.1103/PhysRevC.98.014907} {\bibfield  {journal} {\bibinfo
  {journal} {Phys. Rev. C}\ }\textbf {\bibinfo {volume} {98}},\ \bibinfo
  {pages} {014907} (\bibinfo {year} {2018})},\ \Eprint
  {http://arxiv.org/abs/1807.02948} {arXiv:1807.02948 [hep-ph]} \BibitemShut
  {NoStop}%
\bibitem [{\citenamefont {Vovchenko}(2020)}]{Vovchenko:2020lju}%
  \BibitemOpen
  \bibfield  {author} {\bibinfo {author} {\bibfnamefont {V.}~\bibnamefont
  {Vovchenko}},\ }\href {\doibase 10.1142/S0218301320400029} {\bibfield
  {journal} {\bibinfo  {journal} {Int. J. Mod. Phys. E}\ }\textbf {\bibinfo
  {volume} {29}},\ \bibinfo {pages} {2040002} (\bibinfo {year} {2020})},\
  \Eprint {http://arxiv.org/abs/2004.06331} {arXiv:2004.06331 [nucl-th]}
  \BibitemShut {NoStop}%
\bibitem [{\citenamefont {Steinert}\ and\ \citenamefont
  {Cassing}(2018)}]{Steinert:2018zni}%
  \BibitemOpen
  \bibfield  {author} {\bibinfo {author} {\bibfnamefont {T.}~\bibnamefont
  {Steinert}}\ and\ \bibinfo {author} {\bibfnamefont {W.}~\bibnamefont
  {Cassing}},\ }\href {\doibase 10.1103/PhysRevC.98.014908} {\bibfield
  {journal} {\bibinfo  {journal} {Phys. Rev. C}\ }\textbf {\bibinfo {volume}
  {98}},\ \bibinfo {pages} {014908} (\bibinfo {year} {2018})},\ \Eprint
  {http://arxiv.org/abs/1803.10546} {arXiv:1803.10546 [hep-ph]} \BibitemShut
  {NoStop}%
\bibitem [{\citenamefont {Venugopalan}\ and\ \citenamefont
  {Prakash}(1992)}]{Venugopalan:1992hy}%
  \BibitemOpen
  \bibfield  {author} {\bibinfo {author} {\bibfnamefont {R.}~\bibnamefont
  {Venugopalan}}\ and\ \bibinfo {author} {\bibfnamefont {M.}~\bibnamefont
  {Prakash}},\ }\href {\doibase 10.1016/0375-9474(92)90005-5} {\bibfield
  {journal} {\bibinfo  {journal} {Nucl. Phys.}\ }\textbf {\bibinfo {volume}
  {A546}},\ \bibinfo {pages} {718} (\bibinfo {year} {1992})}\BibitemShut
  {NoStop}%
\bibitem [{\citenamefont {Friman}\ \emph {et~al.}(2015)\citenamefont {Friman},
  \citenamefont {Lo}, \citenamefont {Marczenko}, \citenamefont {Redlich},\ and\
  \citenamefont {Sasaki}}]{Friman:2015zua}%
  \BibitemOpen
  \bibfield  {author} {\bibinfo {author} {\bibfnamefont {B.}~\bibnamefont
  {Friman}}, \bibinfo {author} {\bibfnamefont {P.~M.}\ \bibnamefont {Lo}},
  \bibinfo {author} {\bibfnamefont {M.}~\bibnamefont {Marczenko}}, \bibinfo
  {author} {\bibfnamefont {K.}~\bibnamefont {Redlich}}, \ and\ \bibinfo
  {author} {\bibfnamefont {C.}~\bibnamefont {Sasaki}},\ }\href {\doibase
  10.1103/PhysRevD.92.074003} {\bibfield  {journal} {\bibinfo  {journal} {Phys.
  Rev. D}\ }\textbf {\bibinfo {volume} {92}},\ \bibinfo {pages} {074003}
  (\bibinfo {year} {2015})},\ \Eprint {http://arxiv.org/abs/1507.04183}
  {arXiv:1507.04183 [hep-ph]} \BibitemShut {NoStop}%
\bibitem [{\citenamefont {Vovchenko}\ \emph
  {et~al.}(2018{\natexlab{a}})\citenamefont {Vovchenko}, \citenamefont
  {Motornenko}, \citenamefont {Gorenstein},\ and\ \citenamefont
  {Stoecker}}]{Vovchenko:2017drx}%
  \BibitemOpen
  \bibfield  {author} {\bibinfo {author} {\bibfnamefont {V.}~\bibnamefont
  {Vovchenko}}, \bibinfo {author} {\bibfnamefont {A.}~\bibnamefont
  {Motornenko}}, \bibinfo {author} {\bibfnamefont {M.~I.}\ \bibnamefont
  {Gorenstein}}, \ and\ \bibinfo {author} {\bibfnamefont {H.}~\bibnamefont
  {Stoecker}},\ }\href {\doibase 10.1103/PhysRevC.97.035202} {\bibfield
  {journal} {\bibinfo  {journal} {Phys. Rev. C}\ }\textbf {\bibinfo {volume}
  {97}},\ \bibinfo {pages} {035202} (\bibinfo {year} {2018}{\natexlab{a}})},\
  \Eprint {http://arxiv.org/abs/1710.00693} {arXiv:1710.00693 [nucl-th]}
  \BibitemShut {NoStop}%
\bibitem [{\citenamefont {Lo}\ \emph {et~al.}(2018)\citenamefont {Lo},
  \citenamefont {Friman}, \citenamefont {Redlich},\ and\ \citenamefont
  {Sasaki}}]{Lo:2017lym}%
  \BibitemOpen
  \bibfield  {author} {\bibinfo {author} {\bibfnamefont {P.~M.}\ \bibnamefont
  {Lo}}, \bibinfo {author} {\bibfnamefont {B.}~\bibnamefont {Friman}}, \bibinfo
  {author} {\bibfnamefont {K.}~\bibnamefont {Redlich}}, \ and\ \bibinfo
  {author} {\bibfnamefont {C.}~\bibnamefont {Sasaki}},\ }\href {\doibase
  10.1016/j.physletb.2018.01.016} {\bibfield  {journal} {\bibinfo  {journal}
  {Phys. Lett.}\ }\textbf {\bibinfo {volume} {B778}},\ \bibinfo {pages} {454}
  (\bibinfo {year} {2018})},\ \Eprint {http://arxiv.org/abs/1710.02711}
  {arXiv:1710.02711 [hep-ph]} \BibitemShut {NoStop}%
\bibitem [{\citenamefont {Dash}\ \emph {et~al.}(2018)\citenamefont {Dash},
  \citenamefont {Samanta},\ and\ \citenamefont {Mohanty}}]{Dash:2018can}%
  \BibitemOpen
  \bibfield  {author} {\bibinfo {author} {\bibfnamefont {A.}~\bibnamefont
  {Dash}}, \bibinfo {author} {\bibfnamefont {S.}~\bibnamefont {Samanta}}, \
  and\ \bibinfo {author} {\bibfnamefont {B.}~\bibnamefont {Mohanty}},\ }\href
  {\doibase 10.1103/PhysRevC.97.055208} {\bibfield  {journal} {\bibinfo
  {journal} {Phys. Rev. C}\ }\textbf {\bibinfo {volume} {97}},\ \bibinfo
  {pages} {055208} (\bibinfo {year} {2018})},\ \Eprint
  {http://arxiv.org/abs/1802.04998} {arXiv:1802.04998 [nucl-th]} \BibitemShut
  {NoStop}%
\bibitem [{\citenamefont {Fern\'andez-Ram\'\i{}rez}\ \emph
  {et~al.}(2018)\citenamefont {Fern\'andez-Ram\'\i{}rez}, \citenamefont {Lo},\
  and\ \citenamefont {Petreczky}}]{Fernandez-Ramirez:2018vzu}%
  \BibitemOpen
  \bibfield  {author} {\bibinfo {author} {\bibfnamefont {C.}~\bibnamefont
  {Fern\'andez-Ram\'\i{}rez}}, \bibinfo {author} {\bibfnamefont {P.~M.}\
  \bibnamefont {Lo}}, \ and\ \bibinfo {author} {\bibfnamefont {P.}~\bibnamefont
  {Petreczky}},\ }\href {\doibase 10.1103/PhysRevC.98.044910} {\bibfield
  {journal} {\bibinfo  {journal} {Phys. Rev. C}\ }\textbf {\bibinfo {volume}
  {98}},\ \bibinfo {pages} {044910} (\bibinfo {year} {2018})},\ \Eprint
  {http://arxiv.org/abs/1806.02177} {arXiv:1806.02177 [hep-ph]} \BibitemShut
  {NoStop}%
\bibitem [{\citenamefont {Dash}\ \emph {et~al.}(2019)\citenamefont {Dash},
  \citenamefont {Samanta},\ and\ \citenamefont {Mohanty}}]{Dash:2018mep}%
  \BibitemOpen
  \bibfield  {author} {\bibinfo {author} {\bibfnamefont {A.}~\bibnamefont
  {Dash}}, \bibinfo {author} {\bibfnamefont {S.}~\bibnamefont {Samanta}}, \
  and\ \bibinfo {author} {\bibfnamefont {B.}~\bibnamefont {Mohanty}},\ }\href
  {\doibase 10.1103/PhysRevC.99.044919} {\bibfield  {journal} {\bibinfo
  {journal} {Phys. Rev. C}\ }\textbf {\bibinfo {volume} {99}},\ \bibinfo
  {pages} {044919} (\bibinfo {year} {2019})},\ \Eprint
  {http://arxiv.org/abs/1806.02117} {arXiv:1806.02117 [hep-ph]} \BibitemShut
  {NoStop}%
\bibitem [{\citenamefont {Zyla}\ \emph {et~al.}(2020)\citenamefont {Zyla} \emph
  {et~al.}}]{Zyla:2020zbs}%
  \BibitemOpen
  \bibfield  {author} {\bibinfo {author} {\bibfnamefont {P.}~\bibnamefont
  {Zyla}} \emph {et~al.} (\bibinfo {collaboration} {Particle Data Group}),\
  }\href {\doibase 10.1093/ptep/ptaa104} {\bibfield  {journal} {\bibinfo
  {journal} {PTEP}\ }\textbf {\bibinfo {volume} {2020}},\ \bibinfo {pages}
  {083C01} (\bibinfo {year} {2020})}\BibitemShut {NoStop}%
\bibitem [{\citenamefont {Patrignani}\ \emph {et~al.}(2016)\citenamefont
  {Patrignani} \emph {et~al.}}]{Patrignani:2016xqp}%
  \BibitemOpen
  \bibfield  {author} {\bibinfo {author} {\bibfnamefont {C.}~\bibnamefont
  {Patrignani}} \emph {et~al.} (\bibinfo {collaboration} {Particle Data
  Group}),\ }\href {\doibase 10.1088/1674-1137/40/10/100001} {\bibfield
  {journal} {\bibinfo  {journal} {Chin. Phys.}\ }\textbf {\bibinfo {volume}
  {C40}},\ \bibinfo {pages} {100001} (\bibinfo {year} {2016})}\BibitemShut
  {NoStop}%
\bibitem [{\citenamefont {Capstick}\ and\ \citenamefont
  {Isgur}(1986)}]{Capstick:1986bm}%
  \BibitemOpen
  \bibfield  {author} {\bibinfo {author} {\bibfnamefont {S.}~\bibnamefont
  {Capstick}}\ and\ \bibinfo {author} {\bibfnamefont {N.}~\bibnamefont
  {Isgur}},\ }\bibfield  {booktitle} {\emph {\bibinfo {booktitle}
  {{Proceedings, International Conference on Hadron Spectroscopy: College Park,
  Maryland, April 20-22, 1985}}},\ }\href {\doibase 10.1103/PhysRevD.34.2809,
  10.1063/1.35361} {\bibfield  {journal} {\bibinfo  {journal} {Phys. Rev.}\
  }\textbf {\bibinfo {volume} {D34}},\ \bibinfo {pages} {2809} (\bibinfo {year}
  {1986})},\ \bibinfo {note} {[AIP Conf. Proc.132,267(1985)]}\BibitemShut
  {NoStop}%
\bibitem [{\citenamefont {Godfrey}\ and\ \citenamefont
  {Isgur}(1985)}]{Godfrey:1985xj}%
  \BibitemOpen
  \bibfield  {author} {\bibinfo {author} {\bibfnamefont {S.}~\bibnamefont
  {Godfrey}}\ and\ \bibinfo {author} {\bibfnamefont {N.}~\bibnamefont
  {Isgur}},\ }\href {\doibase 10.1103/PhysRevD.32.189} {\bibfield  {journal}
  {\bibinfo  {journal} {Phys. Rev. D}\ }\textbf {\bibinfo {volume} {32}},\
  \bibinfo {pages} {189} (\bibinfo {year} {1985})}\BibitemShut {NoStop}%
\bibitem [{\citenamefont {Ebert}\ \emph {et~al.}(2009)\citenamefont {Ebert},
  \citenamefont {Faustov},\ and\ \citenamefont {Galkin}}]{Ebert:2009ub}%
  \BibitemOpen
  \bibfield  {author} {\bibinfo {author} {\bibfnamefont {D.}~\bibnamefont
  {Ebert}}, \bibinfo {author} {\bibfnamefont {R.~N.}\ \bibnamefont {Faustov}},
  \ and\ \bibinfo {author} {\bibfnamefont {V.~O.}\ \bibnamefont {Galkin}},\
  }\href {\doibase 10.1103/PhysRevD.79.114029} {\bibfield  {journal} {\bibinfo
  {journal} {Phys. Rev.}\ }\textbf {\bibinfo {volume} {D79}},\ \bibinfo {pages}
  {114029} (\bibinfo {year} {2009})},\ \Eprint {http://arxiv.org/abs/0903.5183}
  {arXiv:0903.5183 [hep-ph]} \BibitemShut {NoStop}%
\bibitem [{\citenamefont {Bollweg}\ \emph {et~al.}(2021)\citenamefont
  {Bollweg}, \citenamefont {Goswami}, \citenamefont {Kaczmarek}, \citenamefont
  {Karsch}, \citenamefont {Mukherjee}, \citenamefont {Petreczky}, \citenamefont
  {Schmidt},\ and\ \citenamefont {Scior}}]{Bollweg:2021vqf}%
  \BibitemOpen
  \bibfield  {author} {\bibinfo {author} {\bibfnamefont {D.}~\bibnamefont
  {Bollweg}}, \bibinfo {author} {\bibfnamefont {J.}~\bibnamefont {Goswami}},
  \bibinfo {author} {\bibfnamefont {O.}~\bibnamefont {Kaczmarek}}, \bibinfo
  {author} {\bibfnamefont {F.}~\bibnamefont {Karsch}}, \bibinfo {author}
  {\bibfnamefont {S.}~\bibnamefont {Mukherjee}}, \bibinfo {author}
  {\bibfnamefont {P.}~\bibnamefont {Petreczky}}, \bibinfo {author}
  {\bibfnamefont {C.}~\bibnamefont {Schmidt}}, \ and\ \bibinfo {author}
  {\bibfnamefont {P.}~\bibnamefont {Scior}},\ }\href@noop {} {\  (\bibinfo
  {year} {2021})},\ \Eprint {http://arxiv.org/abs/2107.10011} {arXiv:2107.10011
  [hep-lat]} \BibitemShut {NoStop}%
\bibitem [{\citenamefont {{The revised Quark Model list used in this work can
  be downloaded at}}()}]{listrepo}%
  \BibitemOpen
  \bibfield  {author} {\bibinfo {author} {\bibnamefont {{The revised Quark
  Model list used in this work can be downloaded at}}},\ }\href@noop {}
  {}\bibinfo {howpublished}
  {\url{http://nsmn1.uh.edu/cratti/decays.html}}\BibitemShut {NoStop}%
\bibitem [{\citenamefont {Taradiy}\ \emph {et~al.}(2019)\citenamefont
  {Taradiy}, \citenamefont {Motornenko}, \citenamefont {Vovchenko},
  \citenamefont {Gorenstein},\ and\ \citenamefont
  {Stoecker}}]{Taradiy:2019taz}%
  \BibitemOpen
  \bibfield  {author} {\bibinfo {author} {\bibfnamefont {K.}~\bibnamefont
  {Taradiy}}, \bibinfo {author} {\bibfnamefont {A.}~\bibnamefont {Motornenko}},
  \bibinfo {author} {\bibfnamefont {V.}~\bibnamefont {Vovchenko}}, \bibinfo
  {author} {\bibfnamefont {M.~I.}\ \bibnamefont {Gorenstein}}, \ and\ \bibinfo
  {author} {\bibfnamefont {H.}~\bibnamefont {Stoecker}},\ }\href {\doibase
  10.1103/PhysRevC.100.065202} {\bibfield  {journal} {\bibinfo  {journal}
  {Phys. Rev. C}\ }\textbf {\bibinfo {volume} {100}},\ \bibinfo {pages}
  {065202} (\bibinfo {year} {2019})},\ \Eprint
  {http://arxiv.org/abs/1904.08259} {arXiv:1904.08259 [hep-ph]} \BibitemShut
  {NoStop}%
\bibitem [{\citenamefont {Vovchenko}\ \emph
  {et~al.}(2018{\natexlab{b}})\citenamefont {Vovchenko}, \citenamefont {Alba},
  \citenamefont {Gorenstein},\ and\ \citenamefont
  {Stoecker}}]{Vovchenko:2017uko}%
  \BibitemOpen
  \bibfield  {author} {\bibinfo {author} {\bibfnamefont {V.}~\bibnamefont
  {Vovchenko}}, \bibinfo {author} {\bibfnamefont {P.}~\bibnamefont {Alba}},
  \bibinfo {author} {\bibfnamefont {M.~I.}\ \bibnamefont {Gorenstein}}, \ and\
  \bibinfo {author} {\bibfnamefont {H.}~\bibnamefont {Stoecker}},\ }\href
  {\doibase 10.1051/epjconf/201817114006} {\bibfield  {journal} {\bibinfo
  {journal} {EPJ Web Conf.}\ }\textbf {\bibinfo {volume} {171}},\ \bibinfo
  {pages} {14006} (\bibinfo {year} {2018}{\natexlab{b}})},\ \Eprint
  {http://arxiv.org/abs/1711.09863} {arXiv:1711.09863 [nucl-th]} \BibitemShut
  {NoStop}%
\bibitem [{\citenamefont {Hagedorn}(1965)}]{Hagedorn:1965st}%
  \BibitemOpen
  \bibfield  {author} {\bibinfo {author} {\bibfnamefont {R.}~\bibnamefont
  {Hagedorn}},\ }\href@noop {} {\bibfield  {journal} {\bibinfo  {journal}
  {Nuovo Cim. Suppl.}\ }\textbf {\bibinfo {volume} {3}},\ \bibinfo {pages}
  {147} (\bibinfo {year} {1965})}\BibitemShut {NoStop}%
\bibitem [{\citenamefont {Lo}\ \emph {et~al.}(2015)\citenamefont {Lo},
  \citenamefont {Marczenko}, \citenamefont {Redlich},\ and\ \citenamefont
  {Sasaki}}]{Lo:2015cca}%
  \BibitemOpen
  \bibfield  {author} {\bibinfo {author} {\bibfnamefont {P.~M.}\ \bibnamefont
  {Lo}}, \bibinfo {author} {\bibfnamefont {M.}~\bibnamefont {Marczenko}},
  \bibinfo {author} {\bibfnamefont {K.}~\bibnamefont {Redlich}}, \ and\
  \bibinfo {author} {\bibfnamefont {C.}~\bibnamefont {Sasaki}},\ }\href
  {\doibase 10.1103/PhysRevC.92.055206} {\bibfield  {journal} {\bibinfo
  {journal} {Phys. Rev. C}\ }\textbf {\bibinfo {volume} {92}},\ \bibinfo
  {pages} {055206} (\bibinfo {year} {2015})},\ \Eprint
  {http://arxiv.org/abs/1507.06398} {arXiv:1507.06398 [nucl-th]} \BibitemShut
  {NoStop}%
\bibitem [{\citenamefont {Vovchenko}\ \emph
  {et~al.}(2018{\natexlab{c}})\citenamefont {Vovchenko}, \citenamefont
  {Gorenstein},\ and\ \citenamefont {Stoecker}}]{Vovchenko:2018fmh}%
  \BibitemOpen
  \bibfield  {author} {\bibinfo {author} {\bibfnamefont {V.}~\bibnamefont
  {Vovchenko}}, \bibinfo {author} {\bibfnamefont {M.~I.}\ \bibnamefont
  {Gorenstein}}, \ and\ \bibinfo {author} {\bibfnamefont {H.}~\bibnamefont
  {Stoecker}},\ }\href {\doibase 10.1103/PhysRevC.98.034906} {\bibfield
  {journal} {\bibinfo  {journal} {Phys. Rev. C}\ }\textbf {\bibinfo {volume}
  {98}},\ \bibinfo {pages} {034906} (\bibinfo {year} {2018}{\natexlab{c}})},\
  \Eprint {http://arxiv.org/abs/1807.02079} {arXiv:1807.02079 [nucl-th]}
  \BibitemShut {NoStop}%
\bibitem [{\citenamefont {Alba}\ \emph {et~al.}(2018)\citenamefont {Alba},
  \citenamefont {Vovchenko}, \citenamefont {Gorenstein},\ and\ \citenamefont
  {Stoecker}}]{Alba:2016hwx}%
  \BibitemOpen
  \bibfield  {author} {\bibinfo {author} {\bibfnamefont {P.}~\bibnamefont
  {Alba}}, \bibinfo {author} {\bibfnamefont {V.}~\bibnamefont {Vovchenko}},
  \bibinfo {author} {\bibfnamefont {M.~I.}\ \bibnamefont {Gorenstein}}, \ and\
  \bibinfo {author} {\bibfnamefont {H.}~\bibnamefont {Stoecker}},\ }\href
  {\doibase 10.1016/j.nuclphysa.2018.03.007} {\bibfield  {journal} {\bibinfo
  {journal} {Nucl. Phys. A}\ }\textbf {\bibinfo {volume} {974}},\ \bibinfo
  {pages} {22} (\bibinfo {year} {2018})},\ \Eprint
  {http://arxiv.org/abs/1606.06542} {arXiv:1606.06542 [hep-ph]} \BibitemShut
  {NoStop}%
\bibitem [{\citenamefont {Vovchenko}\ and\ \citenamefont
  {Koch}(2021)}]{Vovchenko:2020kwg}%
  \BibitemOpen
  \bibfield  {author} {\bibinfo {author} {\bibfnamefont {V.}~\bibnamefont
  {Vovchenko}}\ and\ \bibinfo {author} {\bibfnamefont {V.}~\bibnamefont
  {Koch}},\ }\href {\doibase 10.1103/PhysRevC.103.044903} {\bibfield  {journal}
  {\bibinfo  {journal} {Phys. Rev. C}\ }\textbf {\bibinfo {volume} {103}},\
  \bibinfo {pages} {044903} (\bibinfo {year} {2021})},\ \Eprint
  {http://arxiv.org/abs/2012.09954} {arXiv:2012.09954 [hep-ph]} \BibitemShut
  {NoStop}%
\end{thebibliography}%

\end{document}